\documentclass[12pt]{article}

\usepackage{amsmath,amsfonts,amssymb,amsthm,amsxtra,epsfig,epstopdf,url,array}

\usepackage{graphicx}
\usepackage{color}
\usepackage{latexsym}
\usepackage{bm}

\def\dh#1{\mathop {#1}\limits_{h}}

\def\dh#1{ \mathop{#1}\limits_ h}

\def\dh#1{\mathop {#1}\limits_h}

\def\dvhb#1{\dh#1_{\bar 1}}
\def\dvhb2#1{\dh#1_{\bar 2}}

\def\dh#1{\mathop {#1}\limits_{h}}

\def\dvhb#1{\dh#1_{\bar x}}

\def\dh#1{\mathop {{#1}}\limits_{h}}

\def\dh#1{ \mathop{#1}\limits_ h}

\newcommand{\DD}{\bar{D}}
\newcommand{\Dt}{D_t}
\newcommand{\Dtm}{D_{t^-}}
\newcommand{\Dtp}{D_{t^+}}

\newcommand{\deltau}{{\delta  \over \delta  u} _{(E)}}
\newcommand{\deltauL}{{\delta L \over \delta  u} _{(E)}}

\newtheorem{theorem}{Theorem}[section]
\newtheorem{lemma}[theorem]{Lemma}
\newtheorem{remark}[theorem]{Remark}
\newtheorem{corollary}[theorem]{Corollary}
\newtheorem{proposition}[theorem]{Proposition}
\newtheorem{definition}[theorem]{Definition}

\newtheorem{ex}{Example}[section]
\newenvironment{example}{\begin{ex}\rm}{ \hfill $\Diamond$ \end{ex}
        \vskip4pt}
\newtheorem{ass}{Assumption}[section]

\numberwithin{equation}{section}

\begin{document}

\begin{center}
  {\bf \Large Symmetries and first integrals for variational  ODEs with delay}
\end{center}

\bigskip

\begin{center}
{\large Vladimir  Dorodnitsyn}$^a$,
{\large Roman Kozlov}$^b$,
{\large Sergey  Meleshko}$^c$
\end{center}

\bigskip

\noindent $^a$
Keldysh Institute of Applied Mathematics, Russian Academy of Science, \\
Miusskaya Pl.~4, Moscow, 125047, Russia; \\
{e-mail:  Dorodnitsyn@Keldysh.ru, dorod2007@gmail.com} \\
$^b$ Department of Business and Management Science, Norwegian School
of Economics,
Helleveien 30, 5045, Bergen, Norway;  \\
{e-mail: Roman.Kozlov@nhh.no}  \\
$^c$ School of Mathematics, Institute of Science,
Suranaree University of Technology, 30000, Thailand; \\
{e-mail: sergey@math.sut.ac.th} \\

\bigskip

\bigskip

\begin{center}
{\bf Abstract}
\end{center}
\begin{quotation}
A Lagrangian formalism for variational second-order delay
ordinary differential equations (DODEs) is developed.
The Noether operator identity  for a DODE is established,
which relates the invariance of a Lagrangian function with  the appropriate variational equations and the conserved quantities.
The identity is used to formulate Noether-type theorems that give the first integrals for DODE with symmetries.
Relations between the invariance of the variational second-order DODEs and
the invariance of the Lagrangian functions are also analyzed. Several examples illustrate the theoretical results.
\end{quotation}

\vspace{2pc}

\noindent{\it Keywords}:
delay ordinary differential equations,
variational equations,
Lie point symmetries,
invariance,
Noether's theorem,
first integrals




\section{Introduction}

Lie group analysis has shown an efficient tool for studying ordinary and
partial differential equations since its introduction in the
classical work of Sophus Lie~\cite{bk:Lie[1888],  bk:Lie1924}.
Lie group symmetries of a differential equation transform solutions
into solutions and  can be used to obtain new solutions from known
ones, as well as classify equations into equivalence classes.
Symmetries can also be used to obtain exact analytical solutions that are invariant under
some subgroup of the symmetry group (such solutions are called `group-invariant solutions').
Applications of Lie groups to differential equations are the topic of
many books and
articles~\cite{bk:Ovsiannikov1978, bk:Ibragimov[1983], bk:Olver[1986],
bk:Gaeta1994, bk:HandbookLie, bk:BlumanAnco2002}.
Since  the fundamental work of E.Noether~\cite{Noether1918}
the symmetry group has become a starting point
to obtain first integrals and conservation laws for differential
equations which possess a Lagrangian or Hamiltonian formulation.
The relationships of the symmetry group to conservation laws
for differential equations that do not have a variational statement (and
hence, have no the Lagrangian or Hamiltonian) was developed
in~\cite{bk:AncoBluman1997, bk:BlumanAnco2002}.

Applications of Lie groups of transformations have
been extended to finite-difference, discrete and differential-difference equations and integro-differential equations
\cite{Dorodnitsyn1991, LeviWinternitz1991, QuispelCapelSahadevan,
Dorodnitsyn1993a,  Dorodnitsyn1993,
DorodnitsynKozlovWinternitz2000,  DorKozWin2004,
 LeviWinternitz2005,
bk:Dorodnitsyn[2011],
bk:DorodnitsynKozlov[2011],
Winternitz2011,
bk:Hydon2014, bk:DKapKozWin,
bk:Meleshko[2005],  bk:GrigorievIbragimovKovalevMeleshko2010}.
Exact solutions of delay PDEs were considered in
\cite{Polyanin_Zhurov_2014a, Polyanin_Zhurov_2014b,
Polyanin_Zhurov_2023}.
The symmetry approach was also used to construct numerical schemes,
which preserve qualitative properties of the underlying differential equations
 \cite{bk:ChevDK, bk:DKap1, bk:DKap2,  bk:DKapMel}.

The present article is part of a research project that aims to
extend the application of the group analysis method to variational
delayed ordinary differential equations (DODE). In previous work a Lie group classification of first-order delay ordinary differential equations was presented in ~\cite{bk:DorodnitsynKozlovMeleshkoWinternitz[2018a]},
and linear  first-order delay ordinary differential equations were
considered in ~\cite{bk:DorodnitsynKozlovMeleshkoWinternitz[2018b]}. A Lie group
classification of delay second-order ordinary differential equations
was given in ~\cite{bk:DorodnitsynKozlovMeleshkoWinternitz2021}.

A variational approach was applied to delay differential
equations in \cite{Elsgolts2,Elsgolts3,bk:Elsgolts[1955],Huges1968, Sabbagh1969}
already long time ago, but the symmetries of the variational equations have not been considered yet.
The purpose of this article is to construct the Lagrangian formalism and
Noether-type theorems for delay ordinary differential equations.
For simplicity,  we restrict ourselves to the scalar case and
consider Lagrangian functions with a single delay. Such Lagrangians
provide variational delay ordinary differential equations of second-order (in the sense of the order of differential equations) with two
delays. Thus, an initial value problem for DODEs has to be posed with
{\it two} delays. The Noether theorem for variational DODEs,
developed in the present paper, is  a generalization of Noether's
theorems for variational ordinary differential equations and
variational ordinary difference equations. The DODE analogs of Noether's theorem allow one to find first integrals from Lagrangian functions. If there are sufficiently many first integrals, then they can
be used to express the solutions of a DODE. The relation between
invariance of the  variational  second-order DODEs and the
invariance of the Lagrangian functions, which provide these DODEs,
is also  analyzed. Several examples illustrate the theoretical
results.

The paper is organized as follows.
In the next Section,  we describe second-order delay ordinary differential equations
with two delays.
Section  \ref{section_Lie_point}
describes how to apply Lie point symmetry generators to DODEs.
Delay functionals, their variational equations,  and their invariance
are considered  in Section \ref{section_Variational}.
In Section \ref{local_extremal}, we derive variational equations
for first-order delay functionals  with a single delay,
and obtain second-order DODEs with two delays.
A particular case of such an equation is named the Elsgolts equation.
Definitions of first integrals for a DODE are given in Section \ref{FisrtIttegrals}.
Analogs of Noether's theorem for delay functionals
are formulated and proved in Section   \ref{section_Noether}.
These  analogs are based on the Noether operator identity.
The invariance of the variational DODEs
(in particular, invariance of the Elsgolts equation)
is considered in  Section \ref{section_Invariance}.
In Section \ref{section_Examples},
we demonstrate applications of the developed techniques by several examples.
The final Section  \ref{section_Conclusion}
provides  a short overview of the obtained results.
Some additional results are presented in  Appendices.

\section{Delay ordinary differential equations}

\label{section_DODE}

An initial-value problem for second-order delay ordinary differential equation
with two constant delays can be defined as follows
\begin{subequations}
 \label{DODE0}
\begin{gather}
\label{DODE_EQ}
\ddot{u} ^+ =
F (
\ddot{u},\ddot{u}^-,
\dot{u}^+ , \dot{u} ,\dot{u}^{-},
u^+ , u, u^{-},
t ^+ ,  t , t^-
)  ,
 \qquad
t \in I ,
\\
   \label{delays}
t  ^+  -   t   =  \tau  ,
\qquad
t  -   t ^-  =  \tau  ,
\qquad
\tau = \mbox{const} ,
 \qquad
t \in I ,
\end{gather}
\end{subequations}
where $I \subset  \mathbb{R} $ is some finite or semi{finite} interval.
The independent variable $t$ varies continuously over the entire region, where
Eq.~(\ref{DODE0}) is defined\footnote{In the literature on  DODEs with two delays
it is standard to consider three points
$ t $,   $  t - \tau $ and   $  t - 2 \tau $.
We prefer to use three points  $ t^+  =t  + \tau $,  $ t $ and  $  t^- = t   - \tau   $,
as this choice is more suitable for variational delay equations.
The standard form of a DODE with two delays is given in 
(\ref{DODE1}).}.
We emphasize that the delay parameter takes the same value $\tau$ at all points.
It is convenient to use the right and left shift operators, defined for a function
$ f = f ( t, u) $ as
\begin{equation}  \label{shifts}
f^+ = S_+ (f)   = f(t+\tau, u (t  + \tau)  )  ,
\qquad
f^- = S_{-} (f)= f(t-\tau, u (t  - \tau) ).
\end{equation}
For example,
\[
t^+   = S_+ (t) = t + \tau ,
\qquad
  t^{-}= S_- (t) =  t -  \tau
\]
and
\[
u^+ =   S _+ ( u ) =   u(t + \tau) ,
\qquad
u^{-}  =   S_- (u) =   u(t- \tau) .
\]
The shifts for first- and second-order derivatives are defined similarly:
\[
\dot{u} ^+ =   S _+ ( \dot{u}  ) =   \dot{u}  (t + \tau) ,
\qquad
\dot{u}  ^{-}  =   S_- (  \dot{u} ) =   \dot{u}  (t- \tau) ,
\]
\[
\ddot{u} ^+ =   S _+ ( \ddot{u}  ) =   \ddot{u}  (t + \tau) ,
\qquad
\ddot{u}  ^{-}  =   S_- (  \ddot{u} ) =   \ddot{u}  (t- \tau) .
\]

Notice that the shift operators can be expressed using the differentiation operator
\begin{equation}    \label{one_differentiation}
{\Dt}
=
\frac{\partial {}}{\partial t}
+\dot{u}\frac{\partial { }}{\partial u}
+ \ddot{u} \frac{\partial}{\partial\dot{u}}
+ \cdots
\end{equation}
as
\[
S_+
= {\sum^{}_{s\geq0}{\frac{\tau ^s}{s!}}  D_t^s},
\qquad
S^{-}
= {\sum^{}_{s\geq0}{\frac{(-\tau) ^s}{s!}} D_t^s} .
\]

In order to provide a DODE with two delays
the function $F$ have to satisfy the inequality
\begin{equation}  \label{FDODE}
   \left(  {\partial F  \over  \partial {u}^{-} } \right) ^2
+ \left( {\partial F \over  \partial \dot{u}^{-} } \right) ^2
+ \left( {\partial F \over  \partial \ddot{u}^{-} } \right) ^2
{\not\equiv} 0.
\end{equation}
The initial conditions determine the function $u$ and its derivatives $\dot{u}$ and $\ddot{u}$ on an interval of length $2\tau$.
We denote this interval as $  [  t_0 - \tau , t_0 + \tau  ]  $.

\label{Method_of_steps}

In the literature,
second-order delay ordinary differential equations
with two delays  are usually presented as
\begin{equation}  \label{DODE1}
\ddot{u} (t)  =
F (
\ddot{u} ( t- \tau ) ,\ddot{u} ( t- 2 \tau ) ,
\dot{u} (t)  , \dot{u}  ( t- \tau ),\dot{u} ( t-  2 \tau ),
u (t)  , u ( t- \tau ), u ( t- 2 \tau ) ,
t , \tau )  ,
 \qquad
t \in I ,
\end{equation}
where $I \subset  \mathbb{R} $ is some finite or semi{finite}.
The equation includes a function of both the first and
second-order derivatives at three points $t$, $t-\tau$ and $
t - 2 \tau $, where $\tau$ is the delay parameter. This DODE is of neutral type, because of the presence of the second-order derivatives at the delayed points.
A DODE of the form (\ref{DODE0}) is chosen for the convenience of applications in the variational approach.
DODEs  (\ref{DODE1}) have to be supplemented by initial conditions.
In contrast to the case of ordinary differential equations,
which have initial conditions at a point,
initial conditions for DODE (\ref{DODE1}) are given
on initial interval of length $2 \tau$, e.g.,
\begin{equation}    \label{initialvalues}
u(t)   = \varphi (t),
\qquad
\dot{u}(t)   = \dot{\varphi} (t),
\qquad
\ddot{u}(t)   = \ddot{\varphi} (t),
\qquad
t \in [ t _0  - 2 \tau  , t _0  ]  .
\end{equation}
For simplicity, we assume that the function $\varphi(t)$ is twice differentiable
on the intervals $[t_0 - 2\tau, t_0]$, although this requirement can be relaxed.
One of procedures for solving an DODE either analytically or numerically
is called the {\it method of steps}~\cite{bk:Elsgolts[1955]}.

\section{Lie point symmetries and invariance of DODEs}

\label{section_Lie_point}

Consider an infinitesimal transformation group
\begin{equation}  \label{Group}
t   \rightarrow   t ^*  = f(t, u, a)
\approx
t + \xi  ( t, u) a ,
\qquad
u   \rightarrow  u ^* = g(t,u,a)
\approx
u + \eta  ( t, u) a ,
\end{equation}
where $  a  $  is the group parameter.
Such transformations are represented by a generator in the standard form \cite{bk:Ovsiannikov1978}:
\begin{equation}     \label{operator1}
X
=  {\xi(t,u)}{\partial {} \over \partial t}
+ {\eta(t,u)} {\partial {}   \over \partial u} +  \cdots
\end{equation}

For group analysis of second-order DODE with two delays (\ref{DODE0}),
generators should be prolonged to all variables included in the DODE:
derivatives of $\dot{u} $ and $ \ddot{u}$
and variables at shifted points
$(t^{-},u^{-},\dot{u}^-, \ddot{u}^-) $
and
$(t^{+},u^{+},\dot u^{+}, \ddot{u}^+) $.
This leads to
\begin{multline} \label{operator2}
{X}
={\xi}{\partial \over \partial t}
+ {\eta} {\partial  \over \partial u}
+ {\zeta} _1  {\partial  \over \partial \dot{u}}
+ {\zeta} _2  {\partial  \over \partial \ddot{u}}
\\
+ {\xi}^- {\partial \over \partial t^- }
+ {\eta}^-  {\partial  \over \partial u^- }
+ {\zeta} _1 ^-   {\partial  \over \partial \dot{u}^- }
+ {\zeta} _2  ^- {\partial  \over \partial \ddot{u}^- }
\\
+ {\xi}^+ {\partial \over \partial t^+ }
+ {\eta}^+  {\partial  \over \partial u^+ }
+ {\zeta} _1 ^+   {\partial  \over \partial \dot{u}^+ }
+ {\zeta} _2  ^+ {\partial  \over \partial \ddot{u}^+ }   ,
\end{multline}
where
\[
\xi   =  \xi(t,u)  ,
\qquad
\eta   =  \eta (t,u)  ,
\]
the coefficients
\[
\zeta _1 = \zeta _1  ( t, u ,\dot{u} ) =   {\Dt}    ( \eta ) -  \dot{u}   {\Dt}   ( \xi ) ,
\qquad
\zeta _2 = \zeta _2  ( t, u ,\dot{u} , \ddot{u} ) =  {\Dt}    ( \zeta _1 ) -  \ddot{u}  {\Dt}   ( \xi )
\]
are found according to the standard prolongation formulas
\cite{bk:Ovsiannikov1978,  bk:Ibragimov[1983], bk:Olver[1986]},
and the coefficients
\[
\xi   ^- = S_- ( \xi) =  \xi(t  ^- ,u  ^- )  ,
\qquad
\eta  ^-   = S_-  ( \eta) =  \eta (t ^- ,u ^- )  ,
\]
\[
\xi   ^+ =   S_+ ( \xi )  =   \xi(t  ^+ ,u  ^+ )  ,
\qquad
\eta  ^+   = S_+  (  \eta ) = \eta (t ^+ ,u ^+ )  ,
\]
\[
\zeta _1  ^-  = S_- ( \zeta _1)  =   \zeta _1  ( t ^- , u  ^-  ,\dot{u}  ^-  ) ,
\qquad
\zeta _2  ^- = S_-   ( \zeta _2)  =     \zeta _2  ( t ^- , u  ^- ,\dot{u}  ^- , \ddot{u}  ^- )  ;
\]
\[
\zeta _1  ^+  =   S_+ ( \zeta _1 )    = \zeta _1  ( t ^+ , u  ^+  ,\dot{u}  ^+  ) ,
\qquad
\zeta _2  ^+ =  S_+ ( \zeta _2 )    = \zeta _2  ( t  ^+ , u  ^+ ,\dot{u}  ^+ , \ddot{u}  ^+ )
\]
are obtained by the left and right shift operators  $ S_- $ and  $ S_+  $,
defined in  (\ref{shifts}).

Here and further  $ D _t $ is the  differentiation operator given in  (\ref{one_differentiation})
and  differentiation operators
\begin{equation}      \label{more_differentiation}
{\Dtm}
= \frac{\partial {}}{\partial t^-}
+\dot{u}^- \frac{\partial { }}{\partial u^-}
+ \ddot{u}^{-}\frac{\partial}{\partial\dot{u}^{-}}
+ \cdots  ,
\qquad
{\Dtp}
=\frac{\partial}{\partial t^{+}}
+\dot{u}^{+}\frac{\partial}{\partial u^{+}}
+\ddot{u}^{+}\frac{\partial}{\partial\dot{u}^{+}}
+ \cdots
\end{equation}
correspond to the points $ t^-$ and  $ t^+$,  respectively.
The operator
\begin{equation}
{\DD}
=
{\Dtm}   + {\Dt}  + {\Dtp}
\end{equation}
provides the total derivative.

\begin{remark}
In $\DD$
we separate an operator of total differentiation into three
different operators to preserve its point character. In this case
one can multiply (on the left) them by any functions $\xi$,
$\xi^-$,  and $\xi^+$,  respectively, without losses of
its tangent conditions character (for belonging to higher order
groups or Lie-B\"acklund groups, see \cite{Dorodnitsyn1993a}).
\end{remark}

\section{The invariance  of  delay uniformity in DODEs}

To consider invariant DODEs (\ref{DODE_EQ}) with a constant
delay (\ref{delays}) we require both equations (\ref{DODE_EQ}) and (\ref{delays}) to be invariant together.

In the present paper, we restrict the function $\xi$ to depend on $t$
only: $\xi=\xi(t)$. In this case, one can single out the invariance
condition for system (\ref{DODE_EQ}) and (\ref{delays}), and
consider the invariance of the equations separately. The
infinitesimal criterion for the invariance of an equation
(\ref{DODE_EQ}) becomes
\begin{equation} \label{crit}
\left.
 {X}
(\ddot{u}^+  - F ) \right|_{\ddot{u}^+  =  F,\  t^+-t=t-t^-} =0,
\end{equation}
for the prolonged generators (\ref{operator2}).

{\bf A.} For an arbitrary $t$, the delay parameter $\tau$ is assumed to have the same value to the right and left of $t$:
\begin{equation}
  t^+ -  {t }= t -  {t ^-} = \tau.
\end{equation}
We need this relation to be preserved under group transformations, i.e.,
\[
(t^+) ^* -  t  ^* =t ^* - ( t ^- ) ^* = \tau ^*  .
\]
This leads to the infinitesimal  condition
\[
\xi ( t^+ )  - \xi ( t  ) =
 \xi ( t )  - \xi ( t^-  ).
\]
The latter equation has the following solution:
\begin{equation}\label{regGRID}
\xi (t) = f(t) t   + g(t) ,
\end{equation}
where $f(t)$  and   $g(t)$ are  arbitrary periodic functions with
the period  $\tau$. Notice, that this solution allows the delay
parameter to be changed.

{\bf B.} Moreover, the delay parameter  $ \tau $  has to have the same
value at all points.
Otherwise, the parameter may be changed over time,
remaining equal to the right and left of the point $t$.
Such transformations definitely change the model.
As an example, consider
the transformations corresponding to the generator
\begin{equation*}   
X_{\cos}= t \cos \left( \frac{2\pi t}{\tau}  \right){\frac{\partial}{\partial t}}.
\end{equation*}

Hence, for any two points $t$ and $s$ the delays should have the same
values, i.e.,
\begin{equation}     \label{equal_delay}
  t -  {t ^-} =  s - s ^- =\tau  \qquad  \mbox{for all $t$ and $s$}.
\end{equation}
The latter should be preserved under the group
transformations, i.e.,
\[
  t ^* - ( t ^- ) ^*  =  s ^* - (s ^- ) ^* =\tau ^* .
\]
Although the transformations change the delay parameter from $\tau$ to
$\tau ^*$, the value of the delay parameter is required to be the same at all points.
It leads to the infinitesimal  condition
\[
\xi ( t )  - \xi ( t - \tau  ) = \xi ( s )  - \xi ( s - \tau ).
\]
To satisfy this condition, we arrive at the following result:
\begin{equation}     \label{xiDODE}
\xi (t) = \alpha t   + f(t) ,
\end{equation}
where $\alpha$ is an arbitrary constant and   $f(t)$ is an arbitrary
periodic function with the period  $\tau$.

\begin{figure}[ht]
\centering
\includegraphics[height=120px]{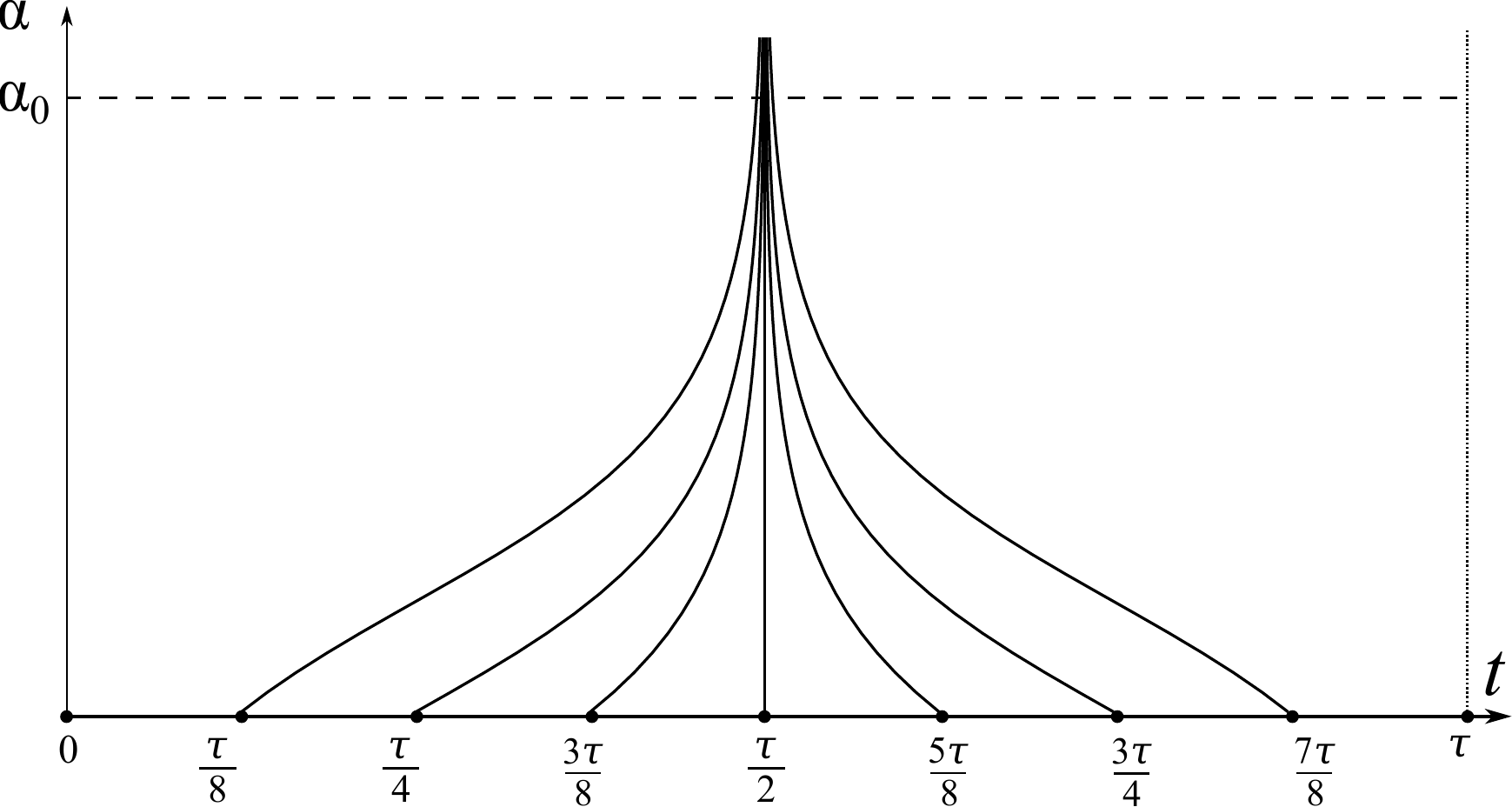}
\caption{Numerical solution for group orbits for points
$t=0,\frac{\tau}{8}, \ldots ,\frac{7\tau}{8},\tau $; $a_0 < \tau$.}
\label{fig:1}
\end{figure}

{\bf C.} The next requirement for a transformation of a delay
is that after the transformation, the time scale should remain unchanged.
Figure \ref{fig:1} shows the {non}linear distortion of the time scale
corresponding to the transformations given by the generator
\begin{equation*}   
 X_{\sin}= \sin \left( \frac{2\pi t}{\tau}  \right) {\frac{\partial}{\partial t}} .
\end{equation*}

Notice that for the transformation with the group parameter $a_0 < \tau$, group orbits actually collapse.
The transformations corresponding to the generator $X_{\sin}$ change the time scale,
and hence the structure of the DODE after transformations.
To keep time homogeneity of transformations,
we have to preserve the following relation for
pairwise distinguished points $t_1,t_2,t_3,t_4$:
\begin{equation}   \label{ratio}
t_1-t_2 = \gamma(t_3-t_4),
\end{equation}
where $\gamma$ is constant.
Applying generator (\ref{xiDODE}), one gets
\begin{equation*}   
 \left( \xi(t_1)- \xi(t_2) -
\gamma(\xi(t_3)-\xi(t_4))\right)|_{(\ref{ratio})}=0.
\end{equation*}
Satisfying  this condition, we obtain the following result:
\begin{equation}
\xi (t) = \alpha t   + \beta   ,\label{cont}
\end{equation}
where $\alpha$ and $\beta$ are arbitrary constants.

Notice that (\ref{cont}) is not required for all results obtained below: some of the requirements can be relaxed.


\section{Invariance of delay functionals}

\label{section_Variational}

Consider a first-order functional with one delay
\begin{equation}      \label{functional_def}
{\cal L} = \int_{a}^{b}
{ L}(t, t^- ,u,u^-,\dot{u},\dot{u}^-)  dt
\end{equation}
and the Lagrangian function $L$ satisfying
\begin{equation}
   \left(  {\partial L   \over  \partial {u} } \right) ^2
+ \left( {\partial L  \over  \partial \dot{u} } \right) ^2
{\not\equiv} 0 ,
\qquad
 \left(  {\partial L   \over  \partial {u}^{-} } \right) ^2
+ \left( {\partial L  \over  \partial \dot{u}^{-} } \right) ^2
{\not\equiv} 0   .
\end{equation}

For a one-parameter group of point transformations  (\ref{Group})
we have
\[  
t ^* = f(t,u,a),
\qquad u ^*=g(t,u,a),
\]
\[  
   (t ^- ) ^* =f^-(t^-,u^-,a),
\qquad (u ^-) ^*     =g^-(t^-,u^-,a) .
\]
Using standard formulas of Lie group analysis
\cite{bk:Ovsiannikov1978,bk:Olver[1986],bk:Ibragimov[1983]},
the transformation has to be prolonged for the derivatives and the differential
\begin{equation*}
dt ^* =  {\Dt}  f(t,u,a) dt .
\end{equation*}

We take an arbitrary interval    $ [ t_1 , t_2] $ such that
   $a   \leq t_1< t_2 \leq b $ and consider the functional
\begin{equation}
{\cal L } _ { [ t_1 , t_2 ] }  = \int_{t_1}^{t_2}{ L}(t, t^- u, u^-, \dot{u}, \dot{u}^-) dt.
\end{equation}
A functional is called invariant with respect to a Lie group of transformation, if
\begin{equation}
  \int_{t_1}^{t_2}{ L}(t, t^- , u,u^-,\dot{u},\dot{u}^-) dt
=\int_{t^*_1}^{t^*_2}{ L}(t ^*,  (t^-)^* ,u ^*, ( u ^-)^*, {\dot{u}} ^*, (\dot{u} ^- )^* ) dt^*.
\end{equation}
Representing the transformed variables and the integration interval
by means of their original values and the group parameter, we have
\begin{equation*}
\int_{t_1}^{t_2}{ L}(t,t^-, u,u^-,\dot{u},\dot{u}^-) dt
=
\int_{t_1}^{t_2}{ L}(f,f^- , g,g^-,\dot{g},\dot{g}^- )   {\Dt}  ( f )   dt.
\end{equation*}
As the interval is arbitrary,  one can omit the integration, and we obtain the
invariance of the elementary action
\begin{equation}\label{Group1}
{ L}(t,t^- , u,u^-,\dot{u},\dot{u}^-) dt
={ L}(f,f^- , g,g^-,\dot{g},\dot{g}^- )  {\Dt}  ( f ) dt.
\end{equation}
Thus, the functional is invariant if and only if the elementary action is invariant.
Differentiating (\ref{Group1}) with respect to the group parameter $a$,
and set $a=0$, we obtain the criterion for invariance of the delay functional.

\begin{theorem}
\label{thm_D1} {\bf (Invariance of a Lagrangian)} The
functional (\ref{functional_def}) is invariant with respect to the group of
transformations  with the generator   (\ref{operator2}) if and only if
\begin{equation}  \label{Group2}
X L + L  {\Dt}  ( \xi  )  = 0 .
\end{equation}
\end{theorem}

In detail, the invariance condition (\ref{Group2}) states\footnote{The paper~\cite{art:FredericoTorres} (see also
\cite{FredericoTorres_2, Second_Noether'stheorem}) contains the
wrong formula for the invariance of a delay Lagrangian.}
\begin{equation}   \label{invariance_functional}
\xi\frac{\partial {L}}{\partial t}
+ \xi^-\frac{\partial {L}}{\partial {t^{-}}}
+ \eta\frac{\partial {L}}{\partial u}
+ \eta^-   \frac{\partial {L}}{\partial  {u^-}}
+ \zeta    _1   \frac{\partial {L}}{\partial \dot u}
+  \zeta^-   _1  \frac{\partial {L}}{\partial \dot u^-}
+ {L}  {\Dt} (\xi)=0  .
\end{equation}

\begin{remark}
In this section, we examine the invariance of the functional  (\ref{functional_def})
without imposing conditions on the invariance of the delay.
To consider  the invariance of a functional and a delay equation,
we need to combine condition (\ref{Group2})  for invariance of the functional
and the condition for invariance of the delay equation.
Invariance of functionals for constant delays (\ref{delays})
is  employed in Section \ref{section_Invariance}.
\end{remark}

We notice that in contrast to the classical ODE case,
there are {\it two} group orbits passing through  the elementary action
${L}(t,t^- , u,u^-,\dot{u},\dot{u}^-)  dt$.
When there are no variables with a delay,
i.e., considering a Lagrangian $ L ( t, u, \dot{u} ) $,
we obtain  the  classical invariance condition
for functionals with first-order Lagrangians
\[
\xi\frac{\partial {L}}{\partial t}
+ \eta\frac{\partial {L}}{\partial u}
+ \zeta   _1   \frac{\partial {L}}{\partial \dot u}
+ {L}  {\Dt} (\xi)=0.
\]

\section{Extremal values of delay functionals
and the locally extremal equation}

\label{local_extremal}

Consider delay functional  (\ref{functional_def}) with a constant delay (\ref{delays}), i.e.,
\begin{subequations}      \label{functional_newdef}
\begin{gather}
{\cal L} = \int_{a}^{b}
{ L}(t, t^- ,u,u^-,\dot{u},\dot{u}^-)  dt
\\
t - t^- = \tau ,
\qquad
\tau = \mbox{const} .
\end{gather}
\end{subequations}


Let the interval    $ [ t_1 , t_2] $ be such that
   $a  \leq  t_1< t_2 \leq   b -\tau $.
We apply slight perturbations of the independent and dependent variables given by
\begin{equation}
\label{eq:09mar.1}
t _{\varepsilon} = t +
\varphi (t)  \varepsilon ,
\qquad
u _{\varepsilon} = u +
\psi (t)  \varepsilon ,
\qquad
t_1   \leq t \leq   t_2  ,
\end{equation}
where
$  \varphi (t) $  and $ \psi  (t) $ are differentiable functions satisfying
\begin{equation}    \label{boundary}
\varphi (s )  =  0
\quad
\mbox{and}
\quad
\psi (s )  =  0
\qquad
\mbox{for}
\qquad
s \in ( - \infty , t_1  ]  \cup [ t_2  , \infty)
\end{equation}
and    $ \varepsilon$ is a small parameter.
Such perturbations produce   variations of the derivative $ \dot{u} $ and the differential $ dt $.

If a function $ u(t) $ provides an extremum of the functional for the given variation,
the perturbed  functional
\begin{equation}
{\cal L}   _{\varepsilon}
= \int_{a}^{b}
 L (t_{\varepsilon}   , t^-_{\varepsilon}    ,u_{\varepsilon}   ,u^-_{\varepsilon}   ,
\dot{u}_{\varepsilon}   ,\dot{u}^-_{\varepsilon}   )  dt_{\varepsilon}
\end{equation}
satisfies the condition
\begin{equation}    \label{extremal_condition}
\left.
{ d  {\cal L}   _{\varepsilon}  \over d  {\varepsilon} }
\right|  _{\varepsilon = 0 }  = 0 .
\end{equation}
Hence, because of conditions (\ref{boundary}), we obtain
\begin{multline*}
\left.
{ d  {\cal L}   _{\varepsilon}  \over d  {\varepsilon} }
\right|  _{\varepsilon = 0 }
=
\int_{t_1}^{t_2}
\left(
  \varphi (t)   \frac{\partial { L}}{\partial t}
+    \psi (t)    \frac{\partial { L}}{\partial u}
+  (    {\Dt}  (  \psi (t)  )     -    \dot{u}  {\Dt}  (  \varphi (t)  )  )
\frac{\partial { L}}{\partial \dot{u}}
+  { L}    {\Dt}  (   \varphi (t)  )
\right) dt
\\
+
 \int_{t_1 + \tau }^{t_2 + \tau }
\left(
\varphi (t^-)   \frac{\partial { L}}{\partial {t^-}}
+  \psi (t^-)   \frac{\partial { L}}{\partial u^-}
 +  (    {\Dtm}  (  \psi (t^-)  )     -    \dot{u}  ^-   {\Dtm}   ( \varphi (t ^- )  )  )
\frac{\partial { L}}{\partial \dot{u}^-}
\right) dt  .
\end{multline*}
Changing the independent variable in the second integral, we get
 \begin{multline*}
\left.
{ d  {\cal L}   _{\varepsilon}  \over d  {\varepsilon} }
\right|  _{\varepsilon = 0 }
=
\int_{t_1}^{t_2}
\left(
 \varphi (t)  \frac{\partial { L}}{\partial t}
+      \psi (t)  \frac{\partial { L}}{\partial u}
+ (    {\Dt}  (  \psi (t)  )     -    \dot{u}    {\Dt}  (  \varphi (t)  )  )
\frac{\partial { L}}{\partial \dot{u}}
+  { L}   {\Dt}  (  \varphi (t)  )
\right) dt
\\
+
 \int_{t_1}^{t_2}
\left(
  \varphi (t)    \frac{\partial { L}^+ }{\partial {t}}
+   \psi (t)   \frac{\partial { L}^+ }{\partial u}
 + (    {\Dt}  (  \psi (t)  )     -    \dot{u}   {\Dt}  (   \varphi (t  )  )  )
\frac{\partial { L}^+ }{\partial \dot{u}}
\right) dt  ,
\end{multline*}
where $  L^+ = { L} (t^+,t, u^+ ,u ,\dot{u}^+ ,\dot{u} ) $
is  the shifted to right the Lagrangian $ L = { L} (t,t^-, u,u^-,\dot{u},\dot{u}^ -)  $.

Employing integration by parts, we have
\begin{multline*}
\left.
{ d  {\cal L}   _{\varepsilon}  \over d  {\varepsilon} }
\right|  _{\varepsilon = 0 }
=
\int_{t_1}^{t_2}
\left\{
 \varphi (t)
\left[
\frac{\partial { L}}{\partial t}
+ \frac{\partial { L}^+ }{\partial {t}}
+ {\DD}
\left(
\dot{u} \frac{\partial { L}}{\partial \dot{u}}
+
 \dot{u}  \frac{\partial { L}^+ }{\partial \dot{u}}
- L
\right)
\right]
\right.
\\
\left.
+
    \psi (t)
\left[
\frac{\partial { L}}{\partial u}
+\frac{\partial { L}^+ }{\partial u}
- {\DD}
\left(
\frac{\partial { L}}{\partial \dot{u}}
+ \frac{\partial { L}^+ }{\partial \dot{u}}
\right)
\right]
\right\} dt ,
\end{multline*}
where boundary terms are excluded due to conditions (\ref{boundary}).

As the interval $[t_1,t_2]$ is arbitrary,  the integrand has to satisfy the equation
\begin{equation}
\label{eq:09mar.2}
\varphi (t)  \left[ { \partial { L} \over \partial t}
+ { \partial { L} ^+ \over \partial t }
+  {\DD} \left( \dot{u}  { \partial { L} \over \partial \dot{u} }
+ \dot{u}  { \partial { L} ^+ \over \partial \dot{u} }
- { L} \right) \right]
+\psi (t)   \left[
 { \partial { L} \over \partial u}
 +  { \partial { L}^+  \over \partial u }
-  {\DD} \left(  { \partial { L} \over \partial \dot{u} }
+
{ \partial { L} ^+ \over \partial  { \dot{u}  } } \right)
 \right] = 0 ,
\end{equation}
which provides a necessary condition for extremum of functional   (\ref{functional_newdef}) for the given variation (\ref{eq:09mar.1}).

For $\varphi=0$ and $\psi\neq 0$  we obtain  the extremal delay equation
\begin{equation}   \label{variational_u}
{\deltauL}
=
\frac{\partial { L}}{\partial u}
+ \frac{\partial { L^+}}{\partial u}
- {\DD}  \left( \frac{\partial { L}}{\partial \dot{u}}
 +   \frac{\partial { L^+}}{\partial \dot{u}} \right)
=0 ,
\end{equation}
which represents the  `vertical variation',
i.e.,  a variational equation for the variation of the dependent variable $u$.
This equation is known since Elsgolts  \cite{bk:Elsgolts[1955]} (see also
 \cite{Elsgolts2, Elsgolts3}).
We call it {\it the Elsgolts equation}.
The corresponding  operator, providing this equation,
\begin{equation}\label{Elsgolts}
{\deltau}
= \frac{\partial {}}{\partial u}
- {\DD} \frac{\partial {}}{\partial \dot{u}}
+ S_+ \left( \frac{\partial {}}{\partial u^-}
-  {\DD} \frac{\partial {}}{\partial
\dot{u}^-} \right),
\end{equation}
is  called {\it the Elsgolts variational derivative}.
Notice that the shift operator $S_+$ acts on {\it all} arguments of functions it is applied to.

The Elsgolts equation operates with ${ L}$ and ${ L}^+$,
while the criterion of the invariance of a Lagrangian includes only $ L$.
We emphasize that, in contrast to the ODE and PDE \cite{Gelfand_Fomin},
the Elsgolts equation generally is not an equation,
on solution of which an invariant Lagrangian
 achieves its extremal value.

For the case $\varphi\neq 0$ and $\psi =0$,
we define a `horizontal variation'
\begin{equation}  \label{variational_t}
{\frac{\delta  { L} }{\delta t}}
  = {  \partial { L} \over \partial t}
+ { \partial { L} ^+ \over \partial t }
+  {\DD} \left( \dot{u}  { \partial { L} \over \partial \dot{u} }
+ \dot{u}  { \partial { L} ^+ \over \partial \dot{u} }
-  { L} \right)
 = 0 .
 \end{equation}
This variational equation corresponds to the variation of the independent variable $t$.
The horizontally variational operator is
\begin{equation}   \label{operator_variational_t}
{ \delta  \over \delta t}
= { \partial    \over \partial t}
+ {\DD} \left( \dot{u}  { \partial       \over \partial \dot{u} } \right)
+ S _+ \left( { \partial       \over \partial t ^-}
+ {\DD} \left( \dot{u} ^- { \partial       \over \partial \dot{u} ^- } \right) \right)
- {\DD}.
\end{equation}

Consider the variation defined by the symmetry (\ref{operator1}). To apply the approach described above, we multiply the coefficients of the generator $X$ by a  function $\varphi(t)$ satisfying (\ref{boundary})
\begin{equation}
t _{\varepsilon} = t +
\varphi(t)\xi (t,u)  \varepsilon ,
\qquad
u _{\varepsilon} = u +
\varphi(t)\eta  (t,u)  \varepsilon.
\end{equation}
Equation (\ref{eq:09mar.2}) leads to the equation
\begin{equation}    \label{localextremal}
\xi \left[ { \partial { L} \over \partial t}
+ { \partial { L} ^+ \over \partial t }
+  {\DD} \left( \dot{u}  { \partial { L} \over \partial \dot{u} }
+ \dot{u}  { \partial { L} ^+ \over \partial \dot{u} }
- { L} \right) \right]
+\eta \left[
 { \partial { L} \over \partial u}
 +  { \partial { L}^+  \over \partial u }
-  {\DD} \left(  { \partial { L} \over \partial \dot{u} }
+
{ \partial { L} ^+ \over \partial  { \dot{u}  } } \right)
 \right] = 0 ,
\end{equation}
which depends {\it explicitly } on $\xi$ and $  \eta$, i.e., on the considered given group, and
it can be rewritten in the form
\begin{equation}         \label{quasi}
\xi   { \delta L \over \delta t }
+  \eta    {\deltauL}
= 0.
\end{equation}

We call it {\it the locally extremal equation}.
Notice that a locally extremal equation is a second-order DODE
containing $ L  $
and the shifted to the right Lagrangian $  L^+  $,
and by virtue of the equation has {\it two delays}.


The locally extremal equation gives the necessary condition for any Lagrangian
to achieve extremal value for variations along orbits of the considered Lie group.
The invariance of a Lagrangian is not needed.
The connection of the
invariance of a Lagrangian with the locally extremal equation is analysed further.

For variations in all possible directions, we obtain the system of equations
\begin{equation}      \label{overdetermined}
\left\{
\begin{array}{l}
{ \displaystyle
{ \delta L \over \delta t }  = 0 ,  } \\
\\
{ \displaystyle
{\deltauL}  = 0 , } \\
\end{array}
\right.
\end{equation}
which can be called the global extremal system.
This system of equations is overdetermined
as discussed in the follow up remark
and can be seen in the examples of Section \ref{section_Examples}.

\begin{remark}   \label{remark_non_equivalence}
It is known that for variational  ordinary  differential equations (without delay)
the vertically and horizontally variational equations
(analogs of    (\ref{variational_u}) and     (\ref{variational_t}))
 are equivalent.
These analogs, namely the Euler–Lagrange equation
\begin{equation}
{ \delta  L \over \delta u}
=
{ \partial  L   \over \partial u}
-    {\Dt}  \left( { \partial    L    \over \partial \dot{u} } \right)  = 0
\end{equation}
and the Du Bois-Reymond equation
\begin{equation}
{ \delta  L \over \delta t}
=
{ \partial    L \over \partial t}
+ {\Dt}  \left( \dot{u}  { \partial  L      \over \partial \dot{u} }   -  L   \right)  =  0 ,
\end{equation}
are determined by the first-order Lagrangians  $ L =  L ( t, u,
\dot{u} )  $. It is not difficult to check that   these equations
are proportional
\begin{equation}     \label{relation}
{ \delta L \over \delta t}
= -    \dot{u}
{ \delta L \over \delta u}   ,
\end{equation}
and,  therefore,  equivalent.

For DODEs the variational  equations (\ref{variational_u}) and
(\ref{variational_t}) are  not equivalent.\footnote{The existence
of two {non}equivalent variational equations for
vertical and horizontal variations was not  recognized in
\cite{art:FredericoTorres} (see also \cite{FredericoTorres_2,
Second_Noether'stheorem}).}


\end{remark}


\begin{remark}
The locally extremal equation    (\ref{localextremal}) depends on the
symmetry coefficients $ \xi (t,u ) $ and  $ \eta (t, u) $.
{Non}equivalence of the vertically variational  equation
(\ref{variational_u}) and  the horizontally    variational  equation
(\ref{variational_t}) forces one to consider different  local
extremal equations for different symmetry operators. If we impose
both equations   (\ref{variational_u})   and (\ref{variational_t}),
it is an overdetermined system   (\ref{overdetermined}).


\end{remark}

\begin{remark}
It should be noted that admitted symmetry algebras might have
symmetries with proportional  (with a {non}constant coefficient of
proportionality) coefficients     $ \xi (t,u) $ and  $ \eta (t, u)
$. Such symmetry operators are called {\it  linearly connected}. For
them,  we obtain equivalent  locally  extremal equations.
Examples illustrating  this property are given in Section  \ref{section_Examples}.
\end{remark}

\section{First integrals of  DODEs}

\label{FisrtIttegrals}

First integrals of second-order ordinary differential equations
\begin{equation*}   
 \ddot{u}  = F (t,u, \dot{u} )
\end{equation*}
have the form
\begin{equation*}  
I( t, u, \dot{u} ),
\end{equation*}
which is not suitable for delay ODEs. Here we generalize this form
of first integrals for ordinary differential equations with delays.

A DODE (\ref{DODE0}) may contain a dependent variable and its derivatives
at three points $t^+ $, $t$ and $t^-$.
Two types of conserved quantities can be introduced: the differential first integral and
the difference first integral.

\begin{definition}   \label{differential_first_integral}
A quantity
\begin{equation}        \label{differential_integral}
I  (   t  ^+,   t, t ^- ,
 u  ^+  ,  u,  u  ^-  ,
 \dot{u}  ^+  ,   \dot{u}  , \dot{u}  ^-     )
\end{equation}
is called {\it a differential first integral} of DODE (\ref{DODE0})
if it holds constant on solutions of the DODE.
\end{definition}

The differential first integral (\ref{differential_integral}) satisfies the equation
\begin{equation}
{\DD} ( I )
=   I _{t ^+ }    +  I _t      +   I _{t ^- }
+ I _{u^+ } \dot{u} ^+    + I _u   \dot{u}    + I _{u  ^- }  \dot{u} ^-
+ I _{ \dot{u} ^+ }    \ddot{u} ^+    + I _{  \dot{u} }    \ddot{u}   + I _{  \dot{u} ^- }    \ddot{u} ^-
= 0 ,
\end{equation}
which should hold for any solution of the considered DODE (\ref{DODE0}).

In addition to the differential first integral, one can define a difference integral.

\begin{definition}   \label{deference_first_integral}
Quantity
\begin{equation}
J  (     t, t ^- ,
      u,  u  ^-  ,
     \dot{u}  , \dot{u}  ^- ,
\ddot{u}  , \ddot{u}  ^-       )
\end{equation}
is called {\it a difference first integral} of DODE (\ref{DODE0}) if
satisfies the equation
\begin{equation}
( S_+   - 1 )   J  = 0
\end{equation}
on the solutions of the DODE.
\end{definition}

We illustrate the definitions of the first integrals by the simple
example.

\begin{example}
The DODE
\[
  \ddot{u} ^+ =     \ddot{u} ^-
\]
has the differential first integral
\[
 I =  \dot{u} ^+ -   \dot{u} ^-
\]
and the difference first integral
\[
  J  =  \ddot{u}   -    \ddot{u} ^-    .
\]
\end{example}

The differential first integral is constant on DODE solutions, while the difference first integral need not be constant on DODE solutions:
it can be a periodic function with period $\tau$, where $\tau$ is the delay parameter.

\section{Noether's identity and Noether-type theorems}

\label{section_Noether}

In this section, we turn to the main results of the paper:
we introduce the Noether operator identity,
which relates the invariance of a Lagrangian (\ref{Group2}),
a locally extremal equation (\ref{localextremal}),
and conserved quantities.

\subsection{Noether's identity}

\begin{lemma}
\label{lem_D3}
{\bf (Noether's identity)}
The following identity holds
\begin{equation}   \label{identity1}
X L+ L {\Dt}( \xi  )
=
\xi { \delta L \over \delta  t }
+
\eta   {\deltauL}
+
{\DD}  ( C^I )
+
( 1 - S_+ )  C^J   ,
\end{equation}
where
\begin{equation}        \label{continuous}
C
=
\xi {L}
 + ( \eta - \dot{u} \xi )   \left( { \partial {L} \over \partial \dot{u} }
  +   { \partial {L} ^+  \over \partial \dot{u} } \right)
\end{equation}
and
\begin{equation}        \label{difference}
P
=
 \xi^-  \frac{\partial {L}}{\partial t^-}
+ \eta^-  \frac{\partial {L}}{\partial u^-}
+  \zeta _1  ^-    \frac{\partial {L}}{\partial \dot{u}^-}   .
\end{equation}
\end{lemma}

\noindent {\it Proof.}
Substituting expressions for the invariance of Lagrangian (\ref{invariance_functional})
and the locally extremal equation (\ref{localextremal}),
which contains variational equations (\ref{variational_u}) and (\ref{variational_t}), in detail equation (\ref{identity1}) is rewritten as
\begin{multline}     \label{detailed-identity}
\xi\frac{\partial {L}}{\partial t}
+ \xi^-\frac{\partial {L}}{\partial t^-}
+\eta\frac{\partial {L}}{\partial u} +
\eta^-\frac{\partial {L}}{\partial {u^-}}
+  ( {\Dt} ( \eta ) - {\dot u}  {\Dt}  (\xi)  )
\frac{\partial {L}}{\partial \dot u}
+  ( {\Dtm}   - {\dot u}^-   {\Dtm}   (\xi ^- )  )
 \frac{\partial {L}}{\partial \dot u^-}
+ {L}   {\Dt} (\xi)
\\
=
\xi \left[ { \partial {L} \over \partial t}
+ { \partial {L} ^+ \over \partial t }
+ {\DD} \left(
\dot{u}  { \partial {L} \over \partial \dot{u} }
+ \dot{u}  { \partial {L} ^+ \over \partial \dot{u} }
-  {L}
\right)
\right]
+
\eta \left[
 { \partial {L} \over \partial u}
 +  { \partial {L}^+  \over \partial u }
- {\DD} \left(
{ \partial {L} \over \partial \dot{u} }
+  { \partial {L} ^+ \over \partial  { \dot{u}  } }
\right)
\right]
\\
+  {\DD}  \left[ \xi {L}
 + ( \eta - \dot{u} \xi )   \left( { \partial {L} \over \partial \dot{u} }
  +   { \partial {L} ^+  \over \partial \dot{u} } \right)
\right]
\\
+
( 1 - S_+)\left( \xi^-  \frac{\partial {L}}{\partial t^-}
+ \eta^-  \frac{\partial {L}}{\partial u^-}
+   ( {\Dtm}  (\eta  ^- )   - {\dot u}^-  {\Dtm}   (\xi ^- )  )
   \frac{\partial {L}}{\partial \dot{u}^-}
\right)   .
\end{multline}
The last identity can be verified directly.
 \hfill $\Box$

\medskip

It should be noted that identity (\ref{identity1})
generalizes Noether's identity for differential equations
\cite{Noether1918, bk:Ibragimov[1983]}
and its difference counterpart
\cite{bk:Dorodnitsyn[2011]}.
The identity yields possibilities
to state various versions of the Noether theorem
for delay differential equations.

\subsection{Analogs of the Noether theorem}

We recall  that invariance  of delay functional
(\ref{Group2}) does not require invariance of the delay equation
(\ref{delays}).

\begin{theorem}  \label{main_Noether}
{\bf (Version of Noether's theorem for DODEs)} Let a delay functional
(\ref{functional_def}) be invariant for the group action generated
by generator (\ref{operator1}) on solutions of the local
extremal equation
\begin{equation}    \label{quasi_extremal}
\xi { \delta L \over \delta  t }
+
\eta
 {\deltauL}
= 0  .
\end{equation}
Then the differential-difference relation
\begin{equation}   \label{dd}
  {\DD}  ( C )
=
( S_+ -1  )  P
\end{equation}
holds on solutions of this equation.
\end{theorem}

\noindent {\it Proof.}
The result follows from identity (\ref{identity1}).
\hfill $\Box$

\medskip

\begin{example}
There are two particular cases in which we get $P\equiv 0$. In these cases, the differential-difference relation (\ref{dd}) gives differential first integrals.

\begin{enumerate}

\item

If a Lagrangian does not depend on  $u$  and $ u^-$,
i.e.,  $  L = L   (   t, t ^- ,     \dot{u}  , \dot{u}  ^-     ) $,
then invariance  with respect to translations of the dependent variable,
represented by the generator
\[
X =  { \partial  \over \partial u}  ,
\]
provides the differential first integral
\[
I
= { \partial {L} \over \partial \dot{u} }
  +   { \partial {L} ^+  \over \partial \dot{u} }  ,
\]
which holds on solutions of the Elsgolts equation (\ref{variational_u}).

\item

If a Lagrangian does not depend on  $t$  and $ t^-$,
i.e.,  $  L = L   (   u, u ^- ,     \dot{u}  , \dot{u}  ^-     ) $,
then invariance  with respect to translations of the independent variable,
represented by the generator
\[
X =  { \partial  \over \partial t}  ,
\]
gives the differential first integral
\[
I
=
 {L}
 - \dot{u}   { \partial {L} \over \partial \dot{u} }
    - \dot{u}    { \partial {L} ^+  \over \partial \dot{u} }  ,
\]
which holds on solutions of the horizontally variational equation (\ref{variational_t}).

\end{enumerate}

\end{example}

Theorem \ref{main_Noether}  has an extension for the {\it divergence invariant} of a  Lagrangian. Such an extension for ODEs was first proposed in \cite{Bessel_Hagen}.

\begin{corollary}    \label{dd_generalization}
Let delay functional  (\ref{functional_def}) satisfy the condition
\begin{equation}   \label{dd_invariance}
X L+ L  {\Dt}  ( \xi  )    =   {\DD}  ( V  )  +   ( 1  -  S_+  )  W
\end{equation}
with some functions
$
V   (   t  ^+,   t, t ^- ,
 u  ^+  ,  u,  u  ^-  ,
 \dot{u}  ^+  ,   \dot{u}  , \dot{u}  ^-     )
$
and
$
W  (     t, t ^- ,
      u,  u  ^-  ,
     \dot{u}  , \dot{u}  ^- ,
\ddot{u}  , \ddot{u}  ^-       )
$
on solutions of locally extremal equation   (\ref{quasi_extremal}).
Then the differential-difference relation
\begin{equation}       \label{dd_divergence}
  {\DD}  ( C - V )
=
( S_+  - 1 )  (  P - W  )
\end{equation}
holds on solutions of this equation.
\end{corollary}

Condition (\ref{dd_invariance})
means the divergent invariance of the Lagrangian.
We call the terms on the right side,
namely
$ {\DD} ( V ) $ and $ ( 1 - S_+ ) W $,
as differential divergence and difference divergence, respectively.

\subsubsection{Homogeneous first integrals.}

For some DODE, differential-difference relation (\ref{dd})
can be converted into a differential first integral or a difference first integral.

\begin{corollary}      \label{proposition_differential}
If there holds
\begin{equation}
( S_+ - 1  )  P  = {\DD}  ( V )
\end{equation}
with some function
$
V   (   t  ^+,   t, t ^- ,
 u  ^+  ,  u,  u  ^-  ,
 \dot{u}  ^+  ,   \dot{u}  , \dot{u}  ^-     )
$,
then the differential-difference relation (\ref{dd}) provides
the differential first integral
\begin{equation}
I = C - V .
\end{equation}
\end{corollary}

\begin{corollary}      \label{proposition_difference}
If there holds
\begin{equation}
{\DD}  ( C )  =  ( S_+  - 1 )   W  ,
\end{equation}
with some function
$
W  (     t, t ^- ,
      u,  u  ^-  ,
     \dot{u}  , \dot{u}  ^- ,
\ddot{u}  , \ddot{u}  ^-       )
$,
then the differential-difference relation (\ref{dd}) provides
the difference  first integral
\begin{equation}
 J =   P  -  W  .
\end{equation}
\end{corollary}

\subsubsection{Special cases of the Noether theorem:
first integrals under additional conditions.}

If the differential-difference relation (\ref{dd})
cannot be converted to differential or difference first integrals, it may be possible to continue with first integrals provided by some {\it condition}.

\begin{proposition}     \label{proposition_difference_constraint}
 {\bf (Locally extremal equation with a difference constraint) }
Let a differential-difference relation (\ref{dd})
hold on solutions of a locally extremal equation
(\ref{quasi_extremal}).
If there holds  the additional condition
\begin{equation}  \label{constraint_difference}
  ( S_+  - 1 )    P  =0,
\end{equation}
then there is the differential first integral
\begin{equation}
I  = C   .
\end{equation}
\end{proposition}

In detail, the additional condition (\ref{constraint_difference}) becomes
\[
 \xi  \frac{\partial {L^+}}{\partial t}
+ \eta  \frac{\partial {L^+}}{\partial u}
+  \zeta _1     \frac{\partial {L^+}}{\partial \dot{u}}
=
 \xi^-  \frac{\partial {L}}{\partial t^-}
+ \eta^-  \frac{\partial {L}}{\partial u^-}
+  \zeta _1  ^-    \frac{\partial {L}}{\partial \dot{u}^-}   .
\]

\begin{proposition}       \label{proposition_differential_constraint}
{\bf (Locally extremal equation with a differential constraint)  }
Let a differential-difference relation (\ref{dd}) hold on
solutions of a locally extremal equation (\ref{quasi_extremal}).
If there holds  the additional condition
\begin{equation}    \label{constraint_differential}
C = \mbox{const} ,
\end{equation}
then there is the difference  first integral
\begin{equation}
J   = P   .
\end{equation}
\end{proposition}

In   Proposition  \ref{proposition_difference_constraint}
the difference first integral stands as an additional condition
and the   differential first integral is obtained as the result.
In   Proposition     \ref{proposition_differential_constraint}
the difference and differential first integral  change their roles.
In general,   constraints  (\ref{constraint_difference})  and
(\ref{constraint_differential}) impose restrictions for a set of
solutions of an DODE.

Other uses of the Noether identity are presented in Appendix \ref{The_other}.

\section{Invariance of the variational equations}

\label{section_Invariance}

The results of this section are valid for generators
(\ref{operator2}) with symmetry coefficients $ \xi (t) $ satisfying
condition (\ref{xiDODE}).

\subsection{The  Elsgolts equation}

First, consider the Elsgolts equation.

\begin{lemma}
For delay functional    (\ref{functional_def})   and generator
(\ref{operator2}),(\ref{xiDODE}) there hold the identity
\begin{equation}     \label{invariant_L_u}
{\deltau}
\left(   X  L   + L    {\Dt} ( {\xi} )   \right)
=
X \left(  {\deltauL}      \right)
+   \left(  \dot{\xi}   + \eta _u       \right)  {\deltauL}   .
\end{equation}
\end{lemma}

\noindent {\it Proof.}
Identity is checked by direct calculation.
\hfill $\Box$

\medskip

This identity allows us to relate
invariance of a Lagrangian and invariance of an Elsgolts equation.

\begin{theorem}    \label{theorem_Invariance_1}
If a Lagrangian is invariant     (\ref{Group2})
with respect to generator   (\ref{operator2}), (\ref{xiDODE}) on solutions of
an Elsgolts equation  (\ref{variational_u}), i.e.,
\begin{equation}    \label{property}
\left.
{\deltau}
\left( X{ L} +{ L}  {\Dt} (\xi) \right)
\right|_{{\deltauL}=0}
=0 ,
\end{equation}
then the Elsgolts equation is also invariant.
\end{theorem}

\noindent {\it Proof.}
From identity   (\ref{invariant_L_u})
and relation  (\ref{variation_of_dd_u})
we obtain
\[
\left.
{\deltau}
\left(X{ L} +{ L}  {\Dt} (\xi)\right)
\right|_{{\deltauL} =0}
=
\left.
X \left(  {\deltauL}  \right)
\right|_{{\deltauL} =0}
\]
which provides the statement of the theorem.
\hfill $\Box$

\medskip

Notice that Theorem \ref{theorem_Invariance_1} matches well
with a similar theorem for ODEs \cite{bk:DorodnitsynKozlov[2011]}.

\begin{ex} Consider Lie algebra $L_{4.13}$
\cite{bk:GonzalezKamranOlver[1992b]}
given by the generators
\begin{equation}
X_{1}=\frac{\partial}{\partial u}, \qquad
X_{2}=\frac{\partial}{\partial t}, \qquad
X_{3}=t\frac{\partial} {\partial t}, \qquad
X_{4}=u\frac{\partial}{\partial u}.
\label{eq:ex1.may29}
\end{equation}
All generators satisfy condition (\ref{cont}), and therefore preserve
the constant delay parameter $\tau$. The complete set of invariants
of this Lie algebra consists of the invariants in the space of the variables  $ \{ t, t^-
, u , u^ - ,   \dot{u} , \dot{u} ^- \} $ and is given by the two invariants
\[
I_{1}=\frac{u-u^{-}}{\dot{u}(t-t^{-})},
\qquad
I_{2}=\frac{\dot{u}^{-}}{\dot{u}} .
\]
Consider the Elsgolts equation with the Lagrangian ${L}={L}(I_{1},I_{2})$.
All generators (\ref{eq:ex1.may29}) satisfy property (\ref{property}).
Thus, the Elsgolts equation admits generators (\ref{eq:ex1.may29}).
Meanwhile, the Lagrangian ${ L}={ L}(I_{1},I_{2})$ does not
admit the scaling generator $X_{3}$.
\end{ex}

\begin{remark}    \label{divergence_Elsgolts}
The statement of Theorem   \ref{theorem_Invariance_1}
is also valid for a divergent invariant Lagrangian,
which satisfy condition   (\ref{dd_invariance}).
This is true because the Elsgolts variational derivative (\ref{Elsgolts})
annihilates the total divergence
\begin{equation}     \label{variation_of_dd_u}
 {\deltau}   (  {\DD} ( V ) + ( S_+ - 1 ) W  )  =  0,
\end{equation}
which can be checked by direct computation.
Notice that here the Elsgolts variational derivative
should be extended to all variables in the expression it acts on, and it has the form
\[  
{\deltau}
= \frac{\partial {}}{\partial u}
- {\DD} \frac{\partial {}}{\partial \dot{u}}
+ {\DD} ^2   \frac{\partial {}}{\partial \ddot{u}}
+ S_+
\left( \frac{\partial {}}{\partial u^-}
-  {\DD} \frac{\partial {}}{\partial \dot{u}^-}
+ {\DD} ^2   \frac{\partial {}}{\partial \ddot{u} ^-}
\right)
+ S_-
\left( \frac{\partial {}}{\partial u^+}
-  {\DD} \frac{\partial {}}{\partial \dot{u}^+}
+ {\DD} ^2   \frac{\partial {}}{\partial \ddot{u} ^+}
\right) .
\]

\end{remark}

\subsection{A horizontally variational equation and a locally extremal equation}

Similar analysis can be performed for a variational equation (\ref{variational_t})
and a locally extremal equation (\ref{localextremal}).

\begin{lemma}
For a delay functional (\ref{functional_def}) and a generator
(\ref{operator2}), (\ref{xiDODE}) the following identity holds
\begin{equation}     \label{invariant_L_t}
{ \delta  \over \delta t }
\left(   X L   + L    {\Dt} ( {\xi} )     \right)
=
X \left(   { \delta  L \over \delta t }      \right)
+  2    \dot{\xi}    { \delta  L \over \delta t }   +  \eta _t  {\deltauL}   .
\end{equation}
\end{lemma}

\noindent {\it Proof.}
The identity is checked by direct calculation.
\hfill $\Box$

\medskip

The identity allows us to relate the invariance of a Lagrangian
and invariance of a horizontally variational equation  (\ref{variational_t}).

\begin{theorem}    \label{theorem_Invariance_2}

Let a Lagrangian be invariant (\ref{Group2})
with respect to a generator  (\ref{operator2}), (\ref{xiDODE})
satisfying the additional condition
\begin{equation}   \label{additional_condition}
\eta _t =0
\end{equation}
on solutions  of the equation (\ref{variational_t}).
Then the horizontally variational equation  (\ref{variational_t}) is also invariant.
\end{theorem}

\noindent {\it Proof.}
The result follows from identity   (\ref{invariant_L_t}).
\hfill $\Box$

\medskip

Invariance of a Lagrangian and invariance of a locally extremal equation  (\ref{localextremal})
can be related as follows.





\begin{lemma}
For a delay functional  (\ref{functional_def}) and a generator
(\ref{operator2}), (\ref{xiDODE}) the following identity holds
\begin{equation}      \label{invariant_L_ut}
\left( \xi  { \delta  \over \delta t } + \eta  {\deltau} \right)
\left(  X   L  + L  {\Dt} ( {\xi} )    \right)
=
X \left(  \xi { \delta  L \over \delta t }   +  \eta {\deltauL}      \right)
+   \dot{\xi}    \left(  \xi { \delta  L \over \delta t }   +  \eta {\deltauL}      \right)  .
\end{equation}
\end{lemma}

\noindent {\it Proof.}
The identify follows from identities      (\ref{invariant_L_u})  and     (\ref{invariant_L_t}).
\hfill $\Box$

\medskip

\begin{theorem}     \label{theorem_Invariance_3}
Let a Lagrangian  be  invariant  (\ref{Group2}) with respect to a generator  (\ref{operator2}), (\ref{xiDODE}) with $ \xi (t)
{\not \equiv} 0 $ and $ \eta (t, u ) {\not \equiv} 0 $ on
solutions  of  locally extremal equation (\ref{quasi_extremal}). Then
the locally extremal equation  (\ref{quasi_extremal}) is also
invariant.
\end{theorem}

\noindent {\it Proof.}
The statement  follows from identity   (\ref{invariant_L_ut}).
\hfill $\Box$

\medskip

\begin{example}    \label{explicitly}
If a Lagrangian does not depend on the independent variable explicitly,
i.e.,
\[
L  =  L  (  u , u ^-   , \dot{u}  , \dot{u}  ^- ) ,
\]
then the horizontally variational equation (\ref{variational_t})
and  the locally extremal equation (\ref{quasi_extremal}) are invariant
with respect to the translation of the independent variable,
represented by the generator
\[
X =  { \partial  \over \partial t } .
\]
It is easy to check that the conditions of Theorems
\ref{theorem_Invariance_2}  and  \ref{theorem_Invariance_3} as well
as condition (\ref{additional_condition}) are satisfied.
\end{example}

\begin{remark}    \label{divergence_horizontal}
As mentioned earlier,  an Elsgolts variational derivative (\ref{Elsgolts})
annihilates a total divergence,
i.e.,  has the property  (\ref{variation_of_dd_u}).
In  contrast to an Elsgolts variational derivative,
a horizontally variational operator  (\ref{operator_variational_t})
does not annihilates a total divergence.
Therefore, the results on invariance of a horizontally variational equation  (\ref{variational_t})
and for a locally extremal equation (\ref{localextremal})
are only formulated for invariant Lagrangians.
\end{remark}

Thus, for symmetries which leave invariant a delay equation
we obtained the following:
a locally extremal equation admits variational symmetries of the delay functional.
For an  Elsgolts  equation and a horizontally variational equation,
which are particular cases of locally extremal equations,
the results are different.
An Elsgolts  equation admits not only variational but also divergence symmetries of the delay functional.
A horizontally variational equation admits variational symmetries of the delay functional, if these symmetries satisfy the additional condition
(\ref{additional_condition}).
The first and second examples of the next section (in subsections \ref{Linear_oscillator_1} and   \ref{DODE_with_time}) illustrate
theorems formulated in this section.

\section{Examples}

\label{section_Examples}

Here we consider the application of the theoretical results
presented in the previous sections.
Another example is given in Appendix \ref{Degenerate_Lagrangian}.

\subsection{Linear oscillator 1}

\label{Linear_oscillator_1}

In this example, we illustrate the basic version of Noether's Theorem \ref{main_Noether},
its corollary, and how to transform differential-difference relations into
differential first integrals.

Consider the Lagrangian function
\begin{equation}         \label{example_1_Lagrangian}
L = \dot{u}  \dot{u} ^-   -      u   u  ^-
\end{equation}
and the symmetries
\begin{equation}
X_1  =  \cos t {  \partial  \over \partial u }  ,
\qquad
X_2  =  \sin t {  \partial  \over \partial u }  ,
\qquad
X_3  =  {  \partial  \over \partial t }  .
\end{equation}

The symmetries $ X_1$  and  $ X_2$  are linearly connected.
For them,  the locally extremal equation  (\ref{quasi_extremal})
is the Elsgolts equation
\begin{equation}     \label{example_1_delta_u}
{\deltauL}
=  -     {u} ^{-}  -   {u} ^{+}    -    \ddot{u} ^{-}  -   \ddot{u} ^{+}
= 0  .
\end{equation}

The symmetry $  X_1  $ satisfies the divergence invariance condition  (\ref{dd_invariance})
\[
{X}_1 L   + L {\Dt} (\xi _1)
=
{\DD} ( -   \sin  t  ^-     {u}    -      \sin  t    {\ } {u} ^{-}    )   .
\]
Therefore, the differential-difference relation  (\ref{dd_divergence}) with
\[
C  _1
=   \cos  t     (   \dot{u} ^{+} +  \dot{u} ^{-} )
+       \sin  t  ^-     {u}   +    \sin  t  {\  }   {u} ^{-}    ,
\qquad
P  _1
=  -  \cos  t  ^-  u  -       \sin  t  ^-   \dot{u}
\]
holds on solutions of equation    (\ref{example_1_delta_u}).
Using Corollary  \ref{proposition_differential}  and
\[
(  S_+ -1 )   P  _1
=  {\DD} (     -       \sin  t  {\ }  {u}  ^+     +        \sin  t  ^-  {u}    )   ,
\]
we find the differential first integral
\begin{equation}
I_1 =  \cos   t     (   \dot{u} ^{+} +  \dot{u} ^{-} )   +   \sin  t  (    {u} ^{+} + {u} ^{-}  )   .
\end{equation}

For the symmetry  $ X_2$,  the divergence invariance condition is
\[
{X} _2  L   + L {\Dt} (\xi _2 )
=
{\DD} (  \cos  t  ^-     {u}   +      \cos  t   {\ }   {u} ^{-}    )   .
\]
The components of the differential-difference relation are
\[
C  _2
=   \sin  t     (   \dot{u} ^{+} +  \dot{u} ^{-} )
-        \cos  t  ^-     {u}  -    \cos  t    {\ }  {u} ^{-}     ,
\qquad
P  _2
=  -  \sin  t  ^-  u  +        \cos  t  ^-   \dot{u}     .
\]
Using
\[
(  S_+ -1 )   P _2
=  {\DD} (        \cos  t   {\ } {u}  ^+    -       \cos  t  ^-  {u}    )  ,
\]
we transform  the differential-difference relation into the differential first integral
\begin{equation}
I_2 =  \sin   t     (   \dot{u} ^{+} +  \dot{u} ^{-} )  -    \cos t  (    {u} ^{+} + {u} ^{-}  )   .
\end{equation}

The differential first integrals  $ I_1 $  and  $ I_2 $ can be used to provide  solutions
of equation (\ref{example_1_delta_u}).
Setting them equal to constants
\[
I_1 = A ,
\qquad
I_2 = B ,
\]
we get
\[
u^+  +  u ^-    = A  \sin t    -   B   \cos t    .
\]

For convenience, we rewrite this relation in the shifted form
\begin{equation}   \label{recursion}
u ( t)  +  u  (t - 2 \tau )     = A  \sin (  t - \tau )     -   B   \cos ( t - \tau )     .
\end{equation}
We start with the initial values
\[
u  (t)   =
\varphi (t)  ,
\qquad
t \in  [ -  2 \tau , 0  ]  .
\]
For simplicity, we assume that the function $\varphi(t)$ has continuous first derivatives on this interval.
The initial conditions give
\[
 A = \dot{\varphi} ( 0 ) +  \dot{\varphi} ( - 2 \tau )  ,
\qquad
B  =   -    {\varphi} ( 0 )   -   {\varphi} ( - 2 \tau )  .
\]

Using   (\ref{recursion}), we obtain
\[
u  (t)
= A  \sin (  t - \tau )       -  B \cos    (  t - \tau )    -  \varphi (t -2 \tau )  ,
\qquad
t \in [ 0 ,  2 \tau ]   ;
\]
\begin{multline*}
u (t)
= A  \sin (  t - \tau )     -  B  \cos    (  t - \tau )   -  u    (t -2 \tau )
\\
= A  \sin (  t - \tau )     -  B  \cos    (  t - \tau )
-  A  \sin (  t - 3 \tau )     +   B  \cos    (  t - 3 \tau )
     +   \varphi (t - 4 \tau )   ,
\qquad
t \in [  2 \tau ,  4 \tau  ]   ;
\end{multline*}
and so on.
By virtue of these relations, one can find  the solution
$ u(t) $, $ t \in [ \tau , \infty )$
recursively starting from the initial data.
In contrast to the method of steps ~\cite{bk:Elsgolts[1955]},
this recursive procedure does not require any integration.

For the symmetry $X_3 $, we get the horizontally variational equation
\begin{equation}     \label{example_1_delta_t}
{ \delta L \over \delta t}
= {\DD}  (  \dot{u}  \dot{u} ^+  + u  u^- )
=  \ddot{u}  \dot{u} ^+    +      \dot{u}  \ddot{u} ^+      +  \dot{u} u^-    + u   \dot{u} ^-
= 0  .
\end{equation}
Notice that equation    (\ref{example_1_delta_u}) is  linear,
while equation    (\ref{example_1_delta_t}) is nonlinear.

The symmetry  $ X_3$ is variational,
i.e.,  it satisfies
\[
{X}_3  L   + L {\Dt} (\xi _3)
=
0  .
\]
It leads to the differential-difference relation (\ref{dd})
with
\[
C  _3
=    -   \dot{u}  \dot{u} ^+    -    u  u^-   ,
\qquad
P  _3
\equiv 0    .
\]
This relation is actually the differential first integral
\begin{equation}
I_3 =  C  _3   =  -    \dot{u}  \dot{u} ^+  -  u  u^-  ,
\end{equation}
which holds on solutions of the  horizontally variational equation   (\ref{example_1_delta_t}).
It is easy to check that the first integral  $ I_3$  holds on solutions of equation  (\ref{example_1_delta_t})
and does not hold on solutions of the Elsgolts equation (\ref{example_1_delta_u}).


This example illustrates the theorems
presented in  Section  \ref{section_Invariance}.
As follows from Theorem  \ref{theorem_Invariance_1}
the Elsgolts equation  (\ref{example_1_delta_u})
admits all symmetries of the Lagrangian:  $ X_1$,   $ X_2$ and   $ X_3$.

We note that the symmetries $X_1$ and $X_2$
give the first integrals $I_1$ and $I_2$ for the Elsholtz equation,
while the symmetry $X_3$ gives the first integral $I_3$
for the horizontally variational equation (\ref{example_1_delta_t}):
the first integral $ I_3$  does not hold on solutions of the Elsgolts equation.

The horizontally variational equation  (\ref{example_1_delta_t})
is invariant with respect to the symmetry  $ X_3 $
as follows from Theorem \ref{theorem_Invariance_2}.
It does not admit symmetries   $ X_1 $ and  $ X_2 $.
Notice that the invariance of both variational equations
(\ref{example_1_delta_u}) and  (\ref{example_1_delta_t})
 for   symmetry   $ X_3 $ can be explained by the form of the Lagrangian,
which does not depend on the independent variable explicitly
(see Example \ref{explicitly}).

Variational equations can admit symmetries,
which are neither variational nor divergence symmetries of the Lagrangians.
For example, both equations (\ref{example_1_delta_u}) and  (\ref{example_1_delta_t})
are invariant with respect to the scaling of the dependent variable,
which is represented  by the generator
\begin{equation}
X_4  =   u  {  \partial  \over \partial u } ,
\end{equation}
while this symmetry is not admitted by the Lagrangian   (\ref{example_1_Lagrangian}):
\[
{X}_4 L   + L {\Dt} (\xi _4)  = 2 L ,
\]
where the right-hand side $2 L$  can not be presented as a divergence.

\subsection{DODE with time-dependent coefficients}

\label{DODE_with_time}

As discussed earlier, Noether's theorem does not require invariance of
delay equations (\ref{delays}).
On the other side,
theorems of Section \ref{section_Invariance}
require invariance of the delay equations in addition to the invariance of a Lagrangian.
To discuss these subtle details
we consider the  delay Lagrangian
\begin{equation}
{L}=   (t^- )  ^{\alpha} {\dot{u}\dot{u}^{-}} ,
\qquad
\alpha  \geq 0
\label{Lag3}
\end{equation}
and  the generators
\begin{equation}   \label{oper3}
X_{1}={\frac{\partial}{\partial u}} ,
\qquad
X_{2}  =   t   ^{\alpha}    {\frac{\partial}{\partial t}} .
\end{equation}
Both generators are variational.
The delay equations (\ref{delays}) are invariant with respect to the generators $ X_1$ and $ X_2$
with  $ \alpha = 0 $ or $ \alpha = 1 $.

For the first symmetry,
we consider the Elsgolts equation
\begin{equation}       \label{example_13_delta_u}
{\deltauL}
=  -   {\alpha}    (t^- )  ^{\alpha -1}  \dot{u}^{-}
-       (t^- )  ^{\alpha}  \ddot{u}^{-}
-   {\alpha}      t ^{\alpha-1}   \dot{u}^{+}
-     t ^{\alpha}   \ddot{u}^{+}
= 0   .
\end{equation}
Because of  $ P_1  \equiv  0 $ the symmetry   provides
 the differential first integral
\[
I _1 = C  _1
=    (t^- )  ^{\alpha}  \dot{u}^{-}   +     t ^{\alpha}   \dot{u}^{+}   .
\]

For the   second  symmetry,   we get the variational  equation
\begin{equation}     \label{example_13_delta_t}
{ \delta L \over \delta t}
=  2   \alpha   t  ^{\alpha-1 } {\dot{u} ^+  \dot{u} }
+   t  ^{\alpha}  \ddot{u} ^+  \dot{u}
+   t  ^{\alpha}  \dot{u} ^+  \ddot{u}
 = 0  .
\end{equation}
As  $ P _2 \equiv  0 $,   the  symmetry  gives
 the differential first integral
\[
I _2 = C   _2
=    t ^{2 \alpha }  \dot{u} ^+ \dot{u}     .
\]

The invariance of a Lagrangian does not require the invariance of a delay equation.
The Noether theorem provides the first integrals  $ I_1 $ and  $ I_2 $,
which hold for any      $ \alpha \geq 0$.
The Elsgolts  equation admits the symmetries of the Lagrangian provided  that
they are also symmetries of the delay equations   (\ref{delays}).
The horizontally variational equation  admits  the variational symmetries of the Lagrangian
provided that $ \eta _t = 0 $ (see Theorem  \ref{theorem_Invariance_2})
and the considered generator  is  also a symmetry of the delay equation.
We obtain that the Elsgolts  equation    (\ref{example_13_delta_u})
and the horizontally variational equation         (\ref{example_13_delta_t})
are invariant for the symmetry $X_1$
and the symmetry $ X_2 $ with the parameter values $ \alpha  = 0 $ and  $ \alpha  = 1 $.

\subsection{Nonlinear DODE}

\label{Nonlinear_DODE}

Consider the  delay Lagrangian
\begin{equation}
{L}=\frac{\dot{u}\dot{u}^{-}}{({u}-{u}^{-})^{2}}
\label{Lag4}
\end{equation}
and  the generators
\begin{equation}   \label{oper4}
X_{1}={\frac{\partial}{\partial u}} ,
\qquad
X_{2}={u\frac{\partial}{\partial u}} ,
\qquad
X_{3}={u^{2}\frac{\partial}{\partial u}} ,
\qquad
X_{4}={\frac{\partial}{\partial t}} .
\end{equation}
All these generators are variational symmetries for the Lagrangian.

For the first three symmetries,
which are linearly connected,
we consider the Elsgolts equation
\begin{equation}       \label{example_3_delta_u}
{\deltauL}
=
-  \frac{2 \dot{u}\dot{u}^{-}}{({u}-{u}^{-})^{3}}
+  \frac{2 \dot{u} ^+ \dot{u}}{({u} ^+ -{u})^{3}}
- {\DD}
\left(
\frac{\dot{u}^{-}}{({u}-{u}^{-})^{2}}
+ \frac{\dot{u}^{+}}{({u}^+-{u})^{2}}
\right)  = 0.
\end{equation}
The generators    $ X_{1} $,  $ X_{2} $ and  $ X_{3} $
provide  the differential-difference relations  (\ref{dd})
with
\[
C  _1
=
\frac{\dot{u}^{-}}{({u}-{u}^{-})^{2}}
+
\frac{\dot{u}^{+}}{({u}^+ -{u})^{2}}  ,
\qquad
P _1
=
 \frac{2 \dot{u}\dot{u}^{-}}{({u}-{u}^{-})^{3}}    ,
\]
\[
C  _2
=
u \left(
\frac{\dot{u}^{-}}{({u}-{u}^{-})^{2}}
+
\frac{\dot{u}^{+}}{({u}^+ -{u})^{2}}
\right)   ,
\qquad
P _2
=
   ({u} + {u}^-)     \frac{2 \dot{u}\dot{u}^{-}}{({u}-{u}^{-})^{3}}
\]
and
\[
C  _3
=
 u^2 \left(
\frac{\dot{u}^{-}}{({u}-{u}^{-})^{2}}
+
\frac{\dot{u}^{+}}{({u}^+ -{u})^{2}}
\right)  ,
\qquad
P _3
=
 2   {u}  {u}^-       \frac{2 \dot{u}\dot{u}^{-}}{({u}-{u}^{-})^{3}}  ,
\]
respectively.

It seems that it is impossible to transform these three differential-difference relations
into differential first integrals.
One can employ Proposition  \ref{proposition_difference_constraint}
and obtain conditional first integrals
\[
I _j = C _ j    , \qquad  j = 1, 2, 3
\]
imposing corresponding difference constraints
\[
( S_+  - 1  )    P _ j   =  0    .
\]
For the symmetries $ X_{1} $,  $ X_{2} $ and  $ X_{3} $
these  difference constraints are
\[
 \frac{2 \dot{u} ^+ \dot{u}}{({u}^+ -{u})^{3}}
=
 \frac{2 \dot{u}\dot{u}^{-}}{({u}-{u}^{-})^{3}}   ,
\]
\[
   ({u}^+  + {u})     \frac{2 \dot{u}^+ \dot{u}}{({u}^+ -{u})^{3}}
=
   ({u} + {u}^-)     \frac{2 \dot{u}\dot{u}^{-}}{({u}-{u}^{-})^{3}}
\]
and
\[
 2   {u} ^+   {u}      \frac{2 \dot{u}^+ \dot{u}}{({u}^+ -{u})^{3}}
=
 2   {u}  {u}^-       \frac{2 \dot{u}\dot{u}^{-}}{({u}-{u}^{-})^{3}} ,
\]
respectively.
Such constraints restrict the set of solutions of the DODE      (\ref{example_3_delta_u}).

For the symmetry  $ X_{4} $,  we get the variational  equation
\begin{equation}     \label{example_3_delta_t}
{ \delta L \over \delta t}
=   {\DD}
\left(
\frac{\dot{u}\dot{u}^{+ }}{({u}^+ -{u})^{2}}
\right)  = 0  .
\end{equation}
The symmetry $ X_{4} $ gives the differential-difference relation with
\[
C _4
=
\frac{\dot{u}\dot{u}^{+ }}{({u}^+ -{u})^{2}}  ,
\qquad
P _4
\equiv 0 .
\]
This means that  we get the differential first integral
\begin{equation*}
I _ 4
= C  _4   =  \frac{\dot{u}\dot{u}^{+ }}{({u}^+ -{u})^{2}}  .
\end{equation*}

\section{Concluding remarks}

\label{section_Conclusion}

The present paper provides  a  Lagrangian formalism for variational delay ordinary differential equations.
For simplicity, we considered first-order Lagrangians with only one delay,
which provide second-order DODEs with two delays.
The appropriate initial value problem for a DODE with two delays was formulated.
The concept of first integrals was generalized for DODEs.
Two types of first integrals were defined:
the differential first integral and the difference first integral.

To consider the invariance of DODE,  not only
the invariance of the DODE itself is required, but also of the equation specifying a delay.
The group transformations must not distort the considered model, which is why
the invariance of delay equation (\ref{delays}) is also required.
We restrict the admitted transformations to those for which the transformation of the independent variable $t$ does not depend on $u$.
It is assumed that the group of transformations can change the delay parameter, but
the transformed  delay parameter must have the same value at all points, which leads to the relation (\ref{xiDODE}).
The invariance of the delay equation restricts the considered
symmetry generators from the general form (\ref{operator1}) to
generators with the coefficient $\xi(t)$ satisfying  condition
(\ref{xiDODE}) or (\ref{cont}).

Furthermore, the invariance condition for a delay functional was established.
It is shown that for directional variation along group orbits, the
extremal problem for a delay functional leads to equation (\ref{quasi}),
called the locally extremal equation. This new  equation  depends
explicitly on the group of transformations considered.
For a `vertical variation', which corresponds to the variation of the dependent
variable, the locally extremal equation becomes  the Elsgolts
equation, named after the mathematician who is the first derived it.
For a `horizontal variation',  one gets another DODE, which
is generally  not equivalent to the Elsgolts equation.

 The Noether-type operator identity relates the
invariance of delay functionals with the locally extremal equation
and the conserved quantities of the variational DODEs. Delay analogs of
the Noether theorem for delay ordinary differential equations of the
second order are presented.  The relation between the invariance of
the variational second-order DODEs and the invariance of the
Lagrangian functions is also analysed. Necessary conditions for the
invariance of the  Elsgolts equation, the invariance of the
horizontally variational equation and the invariance of the local
extremal equation are  established.

Several examples provide applications of the obtained theoretical
results. They show how the Noether-type  theorems can be used for
finding conserved quantities of variational DODEs with symmetries.
If there are sufficiently many first integrals, they can be used  to
express solutions of the  DODEs.

There are many possibilities for generalizing the presented
framework: one can consider higher-order Lagrangians and DODEs, as
well as involve more delays. One can generalize the results to
vector-valued functions.  It is also possible to consider
time-varying delays and delay dependence of solution.

\section*{Dedication}

This paper is dedicated to the memory of Pavel Winternitz in
recognition of his many and diverse contributions to the application of Lie
group methods. The authors  are happy to have been collaborating with him  for
many years.

\section*{Acknowledgments}

The authors thank E.~I.~Kaptsov and {E.~Schulz} for the assistance in the preparation of the paper for publication.


\section*{Appendices}

\appendix

\section{Other treatments of Noether's identity}

\label{The_other}

Noether's identity  (\ref{identity1}) can be used to formulate
other versions of the Noether theorem. It is possible to form
various conditions for the left-hand side or to  form  various DODEs
for the right-hand side and to state corresponding theorems for
conserved quantities (a similar idea was implemented for ODEs in
\cite{{bk:DorodnitsynIbragimov}}). For example, if we aim  to obtain
a differential first integral,
we can keep the locally extremal  equation  (\ref{quasi_extremal}) and
modify the invariance condition (\ref{Group2}).
We rewrite identity  (\ref{identity1}) as
\[
X L+ L {\Dt}( \xi  ) - ( 1 - S_+ )  P = \xi { \delta L \over
\delta  t } + \eta   {\deltauL} + {\DD}  ( C ) .
\]
In detail we get
\begin{multline*}
 \xi \left(\frac{\partial {L}}{\partial t}
+ \frac{\partial {L}^+}{\partial {t}}\right) +
\eta\left(\frac{\partial {L}}{\partial u} +\frac{\partial
{L^+}}{\partial u}\right) +  \zeta _1   \left(\frac{\partial
{L}}{\partial \dot{u}}
 + \frac{\partial {L^+}}{\partial \dot{u}}\right)
 +{L}   {\Dt} (\xi)
\\
= \xi \left[ { \partial {L} \over \partial t} + { \partial {L} ^+
\over \partial t } + {\DD} \left( \dot{u}  { \partial {L} \over
\partial \dot{u} } + \dot{u}  { \partial {L} ^+ \over \partial
\dot{u} } -  {L} \right) \right] + \eta \left[
 { \partial {L} \over \partial u}
 +  { \partial {L}^+  \over \partial u }
- {\DD} \left( { \partial {L} \over \partial \dot{u} } +  { \partial
{L} ^+ \over \partial  { \dot{u}  } } \right) \right]
\\
+  {\DD}  \left[ \xi {L}
 + ( \eta - \dot{u} \xi )   \left( { \partial {L} \over \partial \dot{u} }
  +   { \partial {L} ^+  \over \partial \dot{u} } \right)
\right] .
\end{multline*}
Thus, we arrive at the following results.


\begin{theorem}      \label{theorem_difference_correction_L}
Let a delay functional  (\ref{functional_def}) satisfy  the
condition
\begin{equation}
X L+ L  {\Dt}  ( \xi  )    -  ( 1 - S_+ )  P  =0  ,
\end{equation}
i.e.,
\begin{equation}            \label{difference_correction_L}
\xi \left(\frac{\partial {L}}{\partial t} + \frac{\partial
{L}^+}{\partial {t}}\right) + \eta\left(\frac{\partial {L}}{\partial
u} +\frac{\partial {L^+}}{\partial u}\right) +  \zeta _1
\left(\frac{\partial {L}}{\partial \dot{u}}
 + \frac{\partial {L^+}}{\partial \dot{u}}\right)
 +{L}   {\Dt} (\xi)=0  ,
\end{equation}
on solutions of the locally extremal equation (\ref{quasi_extremal}).
Then the differential first integral
\begin{equation}
I = C
\end{equation}
holds on solutions of this equation.
\end{theorem}

\begin{remark}
Condition  (\ref{difference_correction_L}) was considered in
\cite{art:FredericoTorres}  (see also \cite{FredericoTorres_2,
Second_Noether'stheorem}), where it was stated that an integral of
(\ref{difference_correction_L}) is the invariance condition for the
delay functional (see Theorem 3.4 in  \cite{art:FredericoTorres}).
However,  in the Noether theorem  (see Theorem 3.7 in
\cite{art:FredericoTorres}), the equation for the conserved quantity
was identified incorrectly, namely as the Elsgolts equation
(\ref{variational_u}). Theorem \ref{theorem_difference_correction_L}
presents the correct version of the Noether theorem: the conserved
quantity holds on solutions of the locally extremal equation
(\ref{quasi_extremal}). The locally extremal equation is a linear
combination of the Elsgolts equation  (\ref{variational_u}) and  the
horizontally variational equation   (\ref{variational_t}), which are,
in general case,  not equivalent for delay functionals.
\end{remark}

 \begin{corollary}    \label{corollary_difference_correction_L}
 If a delay functional  (\ref{functional_def}) satisfies the condition
\begin{equation}
 X L+ L {\Dt} ( \xi  )    -  ( 1 - S_+ )  P
 =     {\DD}  ( V ),
\end{equation}
i.e.,
 \begin{equation}             \label{difference_correction_corollary}
  \xi \left(\frac{\partial {L}}{\partial t}
 + \frac{\partial {L}^+}{\partial {t}}\right)
 + \eta\left(\frac{\partial {L}}{\partial u}
 +\frac{\partial {L^+}}{\partial u}\right)
 +  \zeta _1   \left(\frac{\partial {L}}{\partial \dot{u}}
  + \frac{\partial {L^+}}{\partial \dot{u}}\right)
  +{L}  {\Dt} (\xi)
 =     {\DD}  ( V ) ,
 \end{equation}
 with some function
 $
 V  (    t  ^+,   t, t ^- ,
  u  ^+  ,  u,  u  ^-  ,
  \dot{u}  ^+  ,   \dot{u}  , \dot{u}  ^-    )
 $
 on solutions of the locally extremal equation (\ref{quasi_extremal}),
 then the  differential first integral
 \begin{equation}
 I = C  -  V
 \end{equation}
 holds on solutions of this equation.
 \end{corollary}

\subsection{Linear oscillator 2}

\label{Linear_oscillator_2}

This example illustrates
Theorem \ref{theorem_difference_correction_L} and its corollary.
Notice that it can also be analysed in the same manner as
the linear oscillator presented in subsection  \ref{Linear_oscillator_1}.
Consider the following delay Lagrangian
\begin{equation}   \label{Lag2}
{L}
= \frac{(\dot{u}+\dot{u}^{-})^{2}}{2}
- \frac{(u+u^{-})^{2}}{2}  .
\end{equation}
and the symmetries
\begin{equation}
X_1  =  \cos t {  \partial  \over \partial u }  ,
\qquad
X_2  =  \sin t {  \partial  \over \partial u }  ,
\qquad
X_3  =  {  \partial  \over \partial t }  .
\end{equation}

For the  linearly connected symmetries   $X_1$ and $X_2$,   we get the Elsgolts equation
\begin{equation}     \label{example_2_delta_u}
{\deltauL}
=   -  \ddot{u}^{+} -  2\ddot{u} -  \ddot{u}^{-}  - u^{+}   - 2u - u^{-}
=0.
\end{equation}

For the symmetry  $X_1$  the Lagrangian
satisfies the divergence  condition  (\ref{difference_correction_corollary})
with
\[
V =     -  \sin t      (   u^+ + 2 u + u^{-}  )  .
\]
Employing Corollary \ref{corollary_difference_correction_L},
we  get the differential  first integral
\begin{equation}
I_1
=  \cos t  (\dot{u}^{+}+2\dot{u}+\dot{u}^{-})
+ \sin t  ( {u}^{+}+2 {u}+ {u}^{-})
\label{inte2}
\end{equation}
for solutions of the Elsgolts equation    (\ref{example_2_delta_u}).

For the symmetry $ X_2 $ we obtain that the condition (\ref{difference_correction_corollary})
holds with
\[
V  =       \cos t       (   u^+ + 2 u + u^{-}  )       .
\]
Thus,  we get the following differential  first integral
\begin{equation}
I_2
=  \sin t  (\dot{u}^{+}+2\dot{u}+\dot{u}^{-})
-  \cos t  ( {u}^{+}+2 {u}+ {u}^{-})  ,
\end{equation}
which holds on  solutions of equation   (\ref{example_2_delta_u}).

Similar to the previous example
we can use the first integrals  $ I_1 $  and    $ I_2 $
to express the solutions to the initial value problem
for the Elsgolts equation  (\ref{example_2_delta_u}).
Setting  these first integrals equal to constants, we obtain
\[
 {u}^{+}+2 {u}+ {u}^{-}    =  A \sin t   -   B   \cos  t    .
\]
It can be rewritten as
\begin{equation}
 {u} ( t )  + 2   {u}  ( t - \tau )  + {u} ( t - 2 \tau )    = A \sin ( t - \tau )    -   B   \cos  ( t - \tau )
\end{equation}
that can be used to recursively express the solutions to the initial value problem.

The symmetry $ X_3 $ leads  to  the variational equation
\begin{multline}      \label{example_2_delta_t}
{ \delta L \over \delta t}
=
{\DD}
\left(
\dot{u}^{+}  \dot{u}
+ {  3 \over 2 } \dot{u}^2
-  {  1 \over 2 }  (   \dot{u}^{-}) ^2
+
\frac{(u+u^{-})^{2}}{2}
\right)  \\
=
\dot{u} \ddot{u}^{+}
+
\dot{u}^{+}  \ddot{u}
+
3  \dot{u}  \ddot{u}
-
 \dot{u}^{-}  \ddot{u}^{-}
+
(u+u^{-})    ( \dot{u} + \dot{u} ^{-})
=
 0 .
\end{multline}
The symmetry $ X_3 $   satisfies condition  (\ref{difference_correction_L}).
It provides  the following  differential first integral
\begin{equation}
I_3
=
- \dot{u}^{+}  \dot{u}
-  {  3 \over 2 } \dot{u}^2
+   {  1 \over 2 }  (   \dot{u}^{-}) ^2
+-
\frac{(u+u^{-})^{2}}{2}
\end{equation}
for solutions of  equation  (\ref{example_2_delta_t}).

\section{Degenerate Lagrangian function}

\label{Degenerate_Lagrangian}

This example shows the application of many features, which were not
covered by the examples of   Section  \ref{section_Examples}.
In particular,  we show how
to transform differential-difference relations into difference  first
integrals.

Consider the Lagrangian
\begin{equation}
L =    { u - u^- \over t - t^- } \dot{u}  ,
\end{equation}
which leads to the variational derivatives
\[
{\deltauL} =   - { \dot{u} ^+   \over t ^+ - t  } +  { \dot{u} ^-
\over t  - t ^-   }    ,
\]
\[
{ \delta L \over \delta t} =   -  \dot{u}   { \dot{u} - u ^-   \over
(  t  - t^- ) ^2  } +     \dot{u} ^+     {  {u} ^+ - u    \over ( t
^+   - t )  ^2    }    .
\]

Linear Lagrangians do not fit the  generic  framework   presented in
Section~\ref{section_DODE}.
Such Lagrangians do not lead to second-order DODEs. However, all
results concerning conservation properties remain valid.

We consider the symmetries
\begin{equation}
X_1  =  {  \partial  \over \partial u }  , \qquad
X_2  =  t { \partial  \over \partial u }  , \qquad
X_3  =  {  \partial  \over \partial t }  , \qquad
X_4  =  2 t {  \partial  \over \partial t  } + u {  \partial  \over \partial u }  .
\end{equation}

For the symmetries  $ X_1$ and  $ X_2$ we use the Elsgolts equation
\begin{equation}     \label{example_4_delta_u}
 {\deltauL}
=   - { \dot{u} ^+   \over t ^+ - t  } +  { \dot{u} ^-   \over t  -
t ^-   } = 0   .
\end{equation}
In this example, to simplify the equations, we apply the relation
\begin{equation}     \label{help}
t^+ - t = t - t^- =\tau  , \qquad \tau  = \mbox{const} .
\end{equation}
The   Elsgolts equation
(\ref{example_4_delta_u}) becomes
\begin{equation}     \label{example_4_Eq1}
\dot{u} ^+ = \dot{u} ^-   .
\end{equation}

The symmetry $ X_1 $ is variational
\[
X_1 L  + L {\Dt} ( \xi _1 ) =  0 .
\]
It gives the differential-difference relation with
\[
C  _1 = { u - u^- \over t - t^- } ,
\qquad
P _1 = -   { \dot{u} \over t - t^- }   ,
\]
i.e.,
\begin{equation}     \label{dd_original}
{\DD} \left(     { u - u^- \over t - t^- }  \right) =  ( S_+  -1 )
\left(     -   {    \dot{u} \over t - t^- }   \right)    .
\end{equation}
Using  (\ref{help}), we can simplify the densities and the relation
as
\[
\tilde{C}  _1 = { u - u^- } , \qquad \tilde{P} _1 = -   {
\dot{u} }
\]
and
\[
{\DD}  (     { u - u^-  }  ) =  ( S_+  -1 )    (     -   {
\dot{u}  }   )    .
\]

Using Corollary  \ref{proposition_differential} and
\[
( S_+  - 1 )   P _1 = ( S_+  - 1 )     \left(     -   {    \dot{u}
\over t - t^- }   \right) = {\DD}  \left(   -   { u^+  - u \over t -
t^- }  \right)  ,
\]
we can transform  the differential-difference relation
(\ref{dd_original})
  into the differential first integral
\[
I_1  =      { u^+  - u_-  \over t - t^- }     .
\]
Taking into account (\ref{help}), we simplify it as
\[
\tilde{I} _1 = u^+  - u_-   .
\]

Alternatively, we can employ  Corollary
\ref{proposition_difference} and
\[
{\DD} ( C _1   ) = {\DD}
 \left(     { u - u^- \over t - t^- }  \right)
=   ( S_+  - 1 )     \left(        {    \dot{u} _-   \over t - t^- }
\right)
\]
to transform the differential-difference relation  (\ref{dd_original})
into the difference first integral
\[
J _1 =     {    \dot{u}     -     \dot{u} _-   \over t - t^- }     .
\]
Applying  (\ref{help}), we can modify it as
\[
\tilde{J}  _1 =
 \dot{u}     -     \dot{u} _-      .
\]

We observe that the differential first integral   $ \tilde{I} _1$ is
more useful than the difference   first integral  $ \tilde{I} _1 $.
The first integral  $ \tilde{I} _1 $ can be used to express the
solution recursively. No integration is  required.

The symmetry $ X_2 $ is divergence, i.e., it satisfies the condition
(\ref{dd_invariance})
\[
X_2  L   + L  {\Dt} ( \xi_2 ) =   { u - u^- \over t - t^- }    +
\dot{u} =   {\DD} ( u )  +   ( S_+ -1 )  \left(   {  u^- \over t -
t^- }   \right)  .
\]
Notice that both the differential and difference divergence terms are
presented. Corollary \ref{dd_generalization}  provides the
differential-difference relation with
\[
C _2 = t { u - u^- \over t - t^- } -   u
=   {  t^- u - t u^- \over t - t^- }   ,
\qquad
P _2 =   -   {  t^-   \dot{u} \over t - t^- }  -  {  u^- \over t - t^- }
=     {    u^-  - t^- \dot{u}  \over t - t^- }   ,
\]
namely
\[
{\DD}
 \left(     {  t^- u - t u^- \over t - t^- }  \right)
 =  ( S_+  -1 )     \left(       {    u^-  - t^- \dot{u} \over t - t^- }   \right)    .
\]
As the delay is constant,  we can simplify the densities
\[
\tilde{C}  _2 =   {  t^- u - t u^- }   , \qquad
\tilde{P} _2 = {    u^-  - t^- \dot{u} }
\]
and get the relation
\[
{\DD}
 (     {  t^- u - t u^-  }  )
 =  ( S_+  -1 )    (       {    u^-  - t^- \dot{u} }   )    .
\]

For the variational symmetry $ X_3 $, which satisfies the condition
\[
X_3  L + L {\Dt} ( \xi _3 ) =  0 ,
\]
we consider the horizontally variational equation
\begin{equation}      \label{example_4_delta_t}
{ \delta L \over \delta t} =   -  \dot{u}   { \dot{u} - u ^-   \over
(  t  - t^- ) ^2  } +     \dot{u} ^+     {  {u} ^+ - u    \over ( t
^+   - t )  ^2    }  =  0  ,
\end{equation}
which can be simplified
\begin{equation}   \label{example_4_Eq3}
     (    {u} ^+ - u   )     \dot{u} ^+
=       (  \dot{u} - u ^-  )    \dot{u}   .
\end{equation}
The symmetry   $ X_3 $ provides the differential-difference relation
with
\[
C  _3 \equiv  0   , \qquad
P_3 =     \dot{u}      { u  - u ^- \over  ( t - t^- ) ^2  }    .
\]
In other words,  we obtain the difference first integral
\[
J _ 3 =   P _3
=     \dot{u}      { u  - u ^-   \over  ( t - t^- ) ^2  }    .
\]
Employing  (\ref{help}), we simplify it as
\[
\tilde{J} _ 3 =     \dot{u}     (  u  - u ^-  )     .
\]

The symmetry $ X_4 $ is variational:
\[
X_4  L + L {\Dt} ( \xi_4  ) = 0  .
\]
For this symmetry we get the locally extremal equation
\begin{equation}
2 t { \delta L \over \delta t }  + u   {\deltauL} = 2 t \left( -
\dot{u}   { {u} - u ^-   \over (  t  - t^- ) ^2  } +     \dot{u} ^+
{  {u} ^+ - u    \over ( t ^+   - t )  ^2    } \right)
 + u
\left( - { \dot{u} ^+   \over t ^+ - t  } +  { \dot{u} ^-   \over t
- t ^-   } \right)
 = 0  .
\end{equation}
It can be simplified as
\begin{equation}
( 2 t   (     {u} ^+ - u       )   - \tau u )
 \dot{u} ^+
- 2t  (  {u} - u ^-       ) \dot{u} + \tau    u \dot{u} ^-
 = 0  .
\end{equation}

The symmetry gives the differential-difference relation with
\[
C  _4 = u { u - u^- \over t - t^- }      ,
\qquad
P _4 =    2
t^-   \dot{u} {  u - u^-  \over ( t - t^- ) ^2 }  -   {  u^-
\dot{u}   \over t - t^- }    ,
\]
or
\[
{\DD} \left(  u { u - u^- \over t - t^- }    \right) = ( S_+  - 1 )
\left(
 2    t^-   \dot{u} {  u - u^-  \over ( t - t^- ) ^2 }  -   {  u^-   \dot{u}   \over t - t^- }
\right)  .
\]
It can be also simplified to
\[
\tilde{C}  _4 = u  ( u - u^-  )      , \qquad
\tilde{P} _4
= \dot{u}   {    2    t^-    u -  t^-  u^-   - t u^-    \over   t - t^-   }    ,
\]
or
\[
{\DD}
 (  u  ( u - u^-  )    )
= ( S_+  - 1 )  \left(
 \dot{u}   {    2    t^-    u -  t^-  u^-   - t u^-    \over   t - t^-   }
\right)  .
\]


\begin{thebibliography}{10}

\bibitem{bk:Lie[1888]}
S.~Lie.
\newblock {K}lassifikation und {I}ntegration von gew\"ohnlichen
  {D}ifferentialgleichungen zwischen $x,y$,
die eine {G}ruppe von  {T}ransformationen gestatten {I, II}.
\newblock {\em Math. Ann.}, {\bf 32}:213--281, 1888.
\newblock Gesammelte Abhandlungen, vol. 5, B.G. Teubner, Leipzig,
1924, pp.   240--310.


\bibitem{bk:Lie1924}
S.~Lie.
\newblock Gruppenregister.
\newblock {\em Gesammelte Abhandlungen},
{\bf 5} 767--773, 1924.


\bibitem{bk:Ovsiannikov1978}
L.~V.~Ovsiannikov,
\newblock {\em Group Analysis of Differential Equations},
\newblock Nauka, Moscow, 1978.
\newblock {E}nglish translation, {W}.~{F.}~{A}mes, Ed.,
Academic  Press, New York, 1982.


\bibitem{bk:Ibragimov[1983]}
N.~H.~Ibragimov,
\newblock {\em Transformation Groups Applied to Mathematical Physics},
\newblock Nauka, Moscow, 1983.
\newblock {E}nglish translation, Reidel, D., Ed., Dordrecht, 1985.


\bibitem{bk:Olver[1986]}
P.~J.~Olver,
\newblock {\em Applications of {L}ie Groups to Differential Equations},
\newblock Springer-Verlag, New York, 1986.


\bibitem{bk:Gaeta1994}
G.~Gaeta,
\newblock {\em Nonlinear Symmetries and Nonlinear Equations},
\newblock Kluwer, Dordrecht, 1994.


\bibitem{bk:HandbookLie}
N.~H.~Ibragimov, Ed.,
\newblock {\em {CRC} Handbook of {L}ie Group Analysis of Differential   Equations},
volume 1, 2, 3,
\newblock CRC Press, Boca Raton, 1994, 1995, 1996.


\bibitem{bk:BlumanAnco2002}
G.~W.~Bluman and S.~C.~Anco,
\newblock {\em Symmetry and Integration Methods for Differential Equations},
\newblock Springer, New York, 2002.


\bibitem{Noether1918}
E.~Noether.
\newblock Invariante variations problem.
\newblock {\em Nachr. d. {K}\"oniglichen Gesellschaft der Wissenschaften zu
  G\"ottingen, Nachrichten, Mathematisch-Physikalische Klasse Heft 2},
 pages   235--257, 1918.
\newblock English translation: Transport Theory and Statist. Phys.,
{\bf 1} (3) 183-207,    1971, (arXiv:physics/0503066 [physics.hist-ph]).


\bibitem{bk:AncoBluman1997}
S.~C.~Anco and G.~Bluman,
\newblock Direct construction of conservation laws from field equations,
\newblock {\em Phys. Rev. Lett.},
{\bf 78} (3) 2869--2873, 1997.






\bibitem{Dorodnitsyn1991}
V.~A. Dorodnitsyn,
\newblock Transformation groups in net spaces,
\newblock {\em Journal of Soviet Mathematics},
{\bf 55} (1)  1490--1517, 1991.


\bibitem{LeviWinternitz1991}
D.~Levi and P.~Winternitz,
Continuous symmetries of discrete equations,
{\it Phys. Lett.}
{\bf  152},  335--338, 1991


\bibitem{QuispelCapelSahadevan}
G.~R.~W.~Quispel, H.~W.~Capel  and R.~Sahadevan,
\newblock Continuous symmetries of differential-difference equations,
\newblock {\em Phys. Lett. A},
{\bf 170} (5)  379--383, 1992.


\bibitem{Dorodnitsyn1993a}
V.~A.~Dorodnitsyn,
\newblock The finite-difference analogy of Noether's theorem,
\newblock {\em Doklady RAN}, {\bf 328} (6) 678--682, 1993.
\newblock Translation in English in {\it Phys. Dokl.} {\bf 38} (2) 66--68.


\bibitem{Dorodnitsyn1993}
V.~A.~Dorodnitsyn,
\newblock Finite-difference models entirely inheriting symmetry of original
  differential equations,
\newblock In {\em Modern Group Analysis: Advanced Analytical and Computational
  Methods in Mathematical Physics}, volume 191. Kluwer Academic Publishers,
  Boston, 1993.


\bibitem{DorodnitsynKozlovWinternitz2000}
V.~Dorodnitsyn, R.~Kozlov and P.~Winternitz,
\newblock Lie group classification of second-order ordinary difference   equations,
\newblock {\em J. Math. Phys.},
{\bf 41} (1)  480--504, 2000.


\bibitem{DorKozWin2004}
V.~Dorodnitsyn, R.~Kozlov, and P.~Winternitz,
\newblock Continuous symmetries of Lagrangians and exact solutions of discrete   equations,
\newblock {\em J. Math. Phys.},
{\bf 45} (1) 336--359, 2004.


\bibitem{LeviWinternitz2005}
D.~Levi and P.~Winternitz,
\newblock Continuous symmetries of difference equations,
\newblock {\em J. Phys. A: Math. Gen.},
{\bf 39} (2)  R1--R63,   2006.


\bibitem{bk:Dorodnitsyn[2011]}
V.~A.~Dorodnitsyn,
\newblock {\em Applications of Lie Groups to Difference Equations},
\newblock CRC Press, Boca Raton, 2011.


\bibitem{bk:DorodnitsynKozlov[2011]}
V.~A.~Dorodnitsyn and R.~Kozlov,
\newblock {L}agrangian and {H}amiltonian formalism for discrete equations:
  Symmetries and first integrals,
\newblock In D.~Levi  {\it et al.}  Eds.
{\em Symmetries and Integrability of Difference Equations},
pages   7--49,  Cambridge University Press, Cambridge, 2011.
\newblock London Mathematical Society Lecture Notes.


\bibitem{Winternitz2011}
P.~Winternitz,
Symmetry preserving discretization of differential equations
and Lie point symmetries of differential-difference equations,
in  D.~Levi  {\it et al.}  Eds.
{\it Symmetries and Integrability of Difference Equations},
pp 292–341,
Cambridge, Cambridge University Press,    2011.
\newblock London Mathematical Society Lecture Notes.


\bibitem{bk:Meleshko[2005]}
S.~V.~Meleshko,
\newblock {\em Methods for Constructing Exact Solutions of Partial Differential Equations.
Mathematical and Analytical Techniques with Applications to Engineering},
\newblock Springer, New York, 2005.




\bibitem{bk:GrigorievIbragimovKovalevMeleshko2010}
Yu.~N.~Grigoriev, N.~H.~Ibragimov, V.~F.~Kovalev, and S.~V.~Meleshko,
\newblock {\em Symmetries of integro-differential equations and their applications in mechanics and plasma physics},
Lecture Notes in Physics, Vol. 806,
\newblock  Springer, Berlin/Heidelberg, 2010.


\bibitem{bk:Hydon2014}
P.~E.~Hydon,
\newblock {\em Difference Equations by Differential Equation Methods},
\newblock Cambridge University Press, Cambridge, 2014.


\bibitem{bk:DKapKozWin}
V.~A.~Dorodnitsyn, E.~I.~Kaptsov, R.~V.~Kozlov and P.~Winternitz,
The adjoint equation method for constructing first integrals of difference  equations,
{\it J. Phys. A: Math. Theor.}
{\bf 48} (5) 055202, 2015.






\bibitem{bk:ChevDK}
A.~F.~Cheviakov, V.~A.~Dorodnitsyn and E.~I.~Kaptsov,
Invariant conservation law-preserving discretizations of linear and nonlinear
wave equations,
 {\it J. Math. Phys.}  {\bf 61}, 081504, 2020.


\bibitem{bk:DKap1}
V.~A.~Dorodnitsyn and E.~I.~Kaptsov,
Shallow water equations in Lagrangian coordinates:
Symmetries, conservation laws and its preservation in difference models,
{\it Commun. Nonlinear Sci. Numer. Simulat.}   {\bf  89},   105343,  2020.


\bibitem{bk:DKap2}
V.~ A.~Dorodnitsyn and E.~I.~Kaptsov,
Discrete shallow water equations preserving symmetries and conservation laws,
{\it J. Math. Phys.}   {\bf 62} (8)  083508,    2021.



\bibitem{bk:DKapMel}
V.~A.~Dorodnitsyn, E.~I.~Kaptsov, and S.~V.~Meleshko,
Symmetries, conservation laws, invariant solutions
and difference schemes of the one-dimensional Green-Naghdi equations,
{\it J. Nonl. Math.  Phys.},
{\bf  28} (1)  90-107,  2021.


\bibitem{Elsgolts2}
L.~E.~Elsgolts,
\newblock Variational problems with retarded arguments.
\newblock {\em Vestnik Moskovskogo Universiteta. Seriya 1.
Matematika.   Mekhanika},
{\bf 10} 57--62, 1952.
\newblock (in Russian).


\bibitem{Elsgolts3}
L.~E.~Elsgolts,
\newblock Variational problems with retarded arguments,
\newblock {\em Uspekhi Matematicheskikh Nauk},
{\bf 12} (1(73)) 257--258, 1957.
\newblock (in Russian).



\bibitem{bk:Elsgolts[1955]}
L.~E.~Elsgolts,
\newblock {\em Qualative Methods in Mathematical Analysis},
\newblock GITTL, Moscow, 1955.
\newblock Translation: American Mathematical Society, Providence, 1964.

\bibitem{Huges1968}
D.~K.~Hughes,
\newblock Variational and optimal control problems with delayed argument,
\newblock {\em Journal of Optimization Theory and Applications},
{\bf 2} (1)    1–14,
1968.



\bibitem{Sabbagh1969}
L.~D.~Sabbagh,
\newblock Variational problems with lags,
\newblock {\em Journal of Optimization Theory and Applications},
{\bf 3} (1)    34--51,
1969.




\bibitem{bk:DorodnitsynKozlovMeleshkoWinternitz[2018a]}
V.~A.~Dorodnitsyn, R.~Kozlov, S.~V.~Meleshko and P.~Winternitz,
\newblock Lie group classification of first-order delay ordinary differential   equations,
{\it J. Phys. A: Math. Theor.}
{\bf 51} 205202,   2018.


\bibitem{bk:DorodnitsynKozlovMeleshkoWinternitz[2018b]}
V.~A.~Dorodnitsyn, R.~Kozlov, S.~V.~Meleshko and P.~Winternitz,
\newblock Linear or linearizable first-order delay ordinary differential
  equations and their lie point symmetries,
{\it J. Phys. A: Math. Theor.}
{\bf 51}  205203,    2018.


\bibitem{bk:DorodnitsynKozlovMeleshkoWinternitz2021}
V.~A.~Dorodnitsyn, R.~Kozlov, S.~V.~Meleshko and P.~Winternitz,
\newblock Second-order delay ordinary differential equations, their symmetries
  and application to a traffic problem,
{\it J. Phys. A: Math. Theor.}
{\bf 54}   105204,   2021.




\bibitem{Polyanin_Zhurov_2014a}
A. D.  Polyanin and  A. I. Zhurov,
Exact separable solutions of delay reaction-diffusion equations
and other nonlinear partial functional-differential equations,
{\it Commun. Nonlinear Sci. Numer. Simul.},
{\bf 19}  409--16, 2014.


\bibitem{Polyanin_Zhurov_2014b}
A. D.  Polyanin and  A. I. Zhurov,
Nonlinear delay reaction-diffusion equations with varying transfer coefficients:
Exact methods and new solutions,
{\it Appl. Math. Lett.},
{\bf 37}  43--48,  2014.


\bibitem{Polyanin_Zhurov_2023}
A. D.  Polyanin and  V. G. Sorokin,
Reductions and exact solutions of nonlinear wave-type PDEs with proportional
and more complex delays,
{\it Mathematics},
{\bf 11}   516, 2023.







\bibitem{art:FredericoTorres}
G.~S.~F.~Frederico and F.~M.~Torres,
\newblock Noether's symmetry theorem for variational and optimal control
  problems with time delay,
\newblock {\em Numerical Algebra, Control and Optimization (NACO)},
  {\bf 2} (3)  619--630, 2012.


\bibitem{FredericoTorres_2}
G.~S.~F.~Frederico, T.~Odzijewicz and D.~F.~M.~Torres,
Noether's theorem for non-smooth extremals of variational problems with time delay,
{\it  Applicable Analysis},
{\bf 93} (1)   153--170, 2014.


\bibitem{Second_Noether'stheorem}
 A.~B.~Malinowska and T.~Odzijewicz (2016),
Second Noether's theorem with time delay,
 {\it Applicable Analysis}  {\bf 96} (8)  1358--1378, 2017.


\bibitem{Gelfand_Fomin}
I.~M.~Gelfand and S.~V.~Fomin,
{\it  Calculus of variations},
Prentice-Hall,
Englewood Cliffs, N.J.,
1963.




\bibitem{Bessel_Hagen}
E.~Bessel-Hagen,
\newblock \"{U}ber die {E}rhaltungssatze der {E}lektrodynamik,
\newblock {\em Math. Ann.}, {\bf 84} 258--276, 1921.


\bibitem{bk:DorodnitsynIbragimov}
V. Dorodnitsyn and N.Ibragimov.
An extension of the Noether theorem: accompanying equations possessing conservation laws.
{\it Commun. Nonlinear Sci. Numer. Simulat.}
{\bf 19} (2)   328--336,  2014.


\bibitem{bk:GonzalezKamranOlver[1992b]}
A.~Gonzalez-Lopez, N.~Kamran and P.~J.~Olver,
\newblock Lie algebras of vector fields in the real plane,
\newblock {\em Proc. London Math. Soc.},
{\bf 64} 339--368, 1992.



\end{thebibliography}
 \end{document}